\documentclass[structabstract]{aa}  
\usepackage{graphicx}
\usepackage{txfonts}
\usepackage{array}
\usepackage{courier}
\begin{document}
   \title{The VIMOS VLT Deep Survey final data release: a spectroscopic sample of 35\,016 galaxies and AGN out to $z\sim6.7$ selected with $17.5 \leq i_{AB}\leq24.75$  
}

   \author{O. Le F\`evre
          \inst{1}\fnmsep\thanks{Based on data obtained with the European Southern Observatory Very Large
Telescope, Paranal, Chile, under Large Programs 070.A-9007 and 177.A-0837. 
Based on observations obtained with MegaPrime/MegaCam, a joint project of CFHT and CEA/DAPNIA, at the Canada-France-Hawaii Telescope (CFHT) which is operated by the National Research Council (NRC) of Canada, the Institut National des Sciences de l'Univers of the Centre National de la Recherche Scientifique (CNRS) of France, and the University of Hawaii. This work is based in part on data products produced at TERAPIX and the Canadian Astronomy Data Centre as part of the Canada-France-Hawaii Telescope Legacy Survey, a collaborative project of NRC and CNRS.}
\and P. Cassata\inst{1}
\and O. Cucciati\inst{2}
\and B. Garilli\inst{3}
\and O. Ilbert\inst{1}
\and V. Le Brun\inst{1}
\and D. Maccagni\inst{3}
\and C. Moreau\inst{1}
\and M. Scodeggio \inst{3}
\and L. Tresse\inst{1}
\and G. Zamorani \inst{2} 
\and C. Adami \inst{1}
\and S. Arnouts \inst{1}
\and S. Bardelli  \inst{2}
\and M. Bolzonella  \inst{2} 
\and M. Bondi \inst{5}
\and A. Bongiorno \inst{6}
\and D. Bottini \inst{3}
\and A. Cappi    \inst{2}
\and S. Charlot \inst{7}
\and P. Ciliegi    \inst{2}  
\and T. Contini \inst{8}
\and S. de la Torre \inst{9}
\and S. Foucaud \inst{10}
\and P. Franzetti \inst{3}
\and I. Gavignaud \inst{11}
\and L. Guzzo \inst{12}
\and A. Iovino \inst{12}
\and B. Lemaux \inst{1}
\and C. L\'opez-Sanjuan \inst{1,18}
\and H.J. McCracken \inst{7}
\and B. Marano     \inst{6}  
\and C. Marinoni \inst{13}
\and A. Mazure\textdagger \inst{1} 
\and Y. Mellier \inst{7}
\and R. Merighi   \inst{2} 
\and P. Merluzzi \inst{5}
\and S. Paltani \inst{14,15}
\and R. Pell\`o \inst{8}
\and A. Pollo \inst{16,17}
\and L. Pozzetti    \inst{2} 
\and R. Scaramella \inst{4}
\and L. Tasca \inst{1}
\and D. Vergani \inst{2}
\and G. Vettolani \inst{5}
\and A. Zanichelli \inst{5}
\and E. Zucca    \inst{2}
          }

\institute{Aix Marseille Universit\'e, CNRS, LAM - Laboratoire d'Astrophysique de Marseille, 
38 rue F. Joliot-Curie, F-13388, 
Marseille, France 
\and
INAF-Osservatorio Astronomico di Bologna, via Ranzani,1, I-40127,  Bologna, Italy
\and
INAF-IASF, via Bassini 15, I-20133,  Milano, Italy
\and
INAF-Osservatorio Astronomico di Roma, via di Frascati 33, I-00040,  Monte Porzio Catone, Italy
\and
INAF-IRA, via Gobetti,101, I-40129,  Bologna, Italy
\and
Universit\`a di Bologna, Dipartimento di Astronomia, Via Ranzani 1, I-40127, 
Bologna, Italy
\and
Institut d'Astrophysique de Paris, UMR7095 CNRS,
Universit\'e Pierre et Marie Curie, 98 bis Boulevard Arago, 75014
Paris, France
\and
Institut de Recherche en Astrophysique et Plan\'etologie, 14, avenue E. Belin, F31400 
Toulouse, France
\and
SUPA, Institute for Astronomy, University of Edinburgh, Royal Observatory,
Blackford Hill, Edinburgh EH9 3HJ, 
Edinburgh, UK
\and
Department of Earth Sciences, National Taiwan Normal University, 
No. 88, Section 4, Tingzhou Road, Wenshan District, Taipei 11677, Taiwan
Taipei, Taiwan
\and
Departamento de Ciencias Fisicas, Universidad Andres Bello, Santiago, Chile
\and
INAF - Osservatorio Astronomico di Brera, via E. Bianchi 46, Merate / Via Brera 28, 
Milano, Italy
\and
Aix-Marseille Universit\'e, CNRS, Centre de Physique Th\'eorique,  F-13288 
Marseille, France
\and
Integral Science Data Centre, Gen\`eve, ch. d'\'Ecogia 16, CH-1290 Versoix
Switzerland
\and
Geneva Observatory, ch. des Maillettes 51, CH-1290 Sauverny, 
Gen\`eve, Switzerland
\and
Astronomical Observatory of the Jagiellonian University, ul Orla 171, 30-244 
Krak{\'o}w, Poland
\and
National Centre for Nuclear Research, Hoza 69, 00-681, Warszawa, Poland
\and
Centro de Estudios de F\'isica del Cosmos de Arag\'on, Teruel, Spain
 \\ \\
             \email{olivier.lefevre@lam.fr}
             }

   \date{Received 30 June 2013; accepted 22 August 2013}

 
  \abstract
   {Deep representative surveys of galaxies at different epochs 
    are needed to make progress in understanding galaxy evolution.}
   {We describe the completed VIMOS VLT Deep Survey, and  the final data release of 35\,016
galaxies and  type-I AGN with measured spectroscopic redshifts covering all epochs up to redshift $z\sim6.7$,
in areas from 0.142 to 8.7 square degrees, and volumes from $0.5\times10^6$ to $2\times10^7$h$^{-3}$Mpc$^3$. }
   {We have selected samples of galaxies based solely on their i-band magnitude reaching $i_{AB}=24.75$. 
    Spectra have been obtained with VIMOS on the ESO-VLT 
    integrating 0.75h, 4.5h and 18h for the Wide, Deep, and Ultra-Deep nested surveys, respectively.
    We demonstrate that any 'redshift desert' can be crossed successfully using
    spectra covering $3650 \leq \lambda \leq 9350$\AA. 
    A total of 1263 galaxies have been re-observed independently within the VVDS, and from the VIPERS and MASSIV surveys.
    They are used to establish the redshift measurements reliability,
    to assess completeness in the VVDS samples, and to provide a weighting scheme taking
    into account the survey selection function.     
    We describe the main properties of the VVDS samples, 
    and the VVDS is compared to other spectroscopic surveys in the literature. 
}
   {In total we have obtained spectroscopic redshifts for 34\,594 galaxies, 422 type-I AGN, and 12\,430 Galactic
    stars. The survey has enabled to identify galaxies up to very high redshifts
    with 4669 redshifts in $1 \leq z_{spec} \leq 2$, 561 in $2 \leq z_{spec} \leq 3$ and 468 with  $z_{spec} > 3$,
    and specific populations like Lyman-$\alpha$ emitters have been identified out to $z=6.62$. We show that the VVDS occupies a unique
    place in the parameter space defined by area, depth, redshift coverage, and number of spectra.
     }
   {The VIMOS VLT Deep Survey provides a comprehensive survey of the distant universe, covering
   all epochs since $z\sim6$, or more than 12 Gyr of cosmic time, with a uniform selection, the largest such sample 
   to date. A wealth of science results derived from the VVDS have shed new light on the evolution
   of galaxies and AGN, and their distribution in space, over this large cosmic time.  
   The VVDS further demonstrates that large deep spectroscopic redshift surveys 
   over all these epochs in the distant Universe are a key tool to observational cosmology.
   To enhance the Legacy value of the survey, a final public release of the complete VVDS spectroscopic redshift sample
   is available at \texttt{http://cesam.lam.fr/vvds}.}


\authorrunning{Le F\`evre, O. et al.}
\titlerunning{VVDS final data release: 35016 galaxies and AGN out to $z\sim6.7$}

   \maketitle
%

\section{Introduction}

A fundamental goal of observational cosmology is to understand the formation and evolution of galaxies 
and their distribution in space, in the cosmological framework of
an evolving universe. Progress in this field closely follows progress
in assembling large samples of galaxies from deep surveys going at ever 
earlier epochs towards the time when the first objects have been forming.

In a now standard observing strategy, galaxies are first identified using deep imaging
surveys, then their redshifts are measured and enable a global statistical description
of the galaxy population. This may then allow to target specific sub-populations
for higher spatial and/or spectral resolution, e.g. with integral field spectroscopy, or more extended 
wavelength coverage. The knowledge of the spectroscopic redshift of a galaxy is a key information, 
giving not only the distance in relation to cosmic time, but also the position in space 
in relation to the local network of large scale structures. From the same observation,
spectra also provide a wealth of information on the stellar, AGN, dust and gas content of each 
galaxy. 

Spectroscopic redshifts 
surveys have developed from these simple ideas, and have been strongly pushing large telescopes
and instrumentation development. The statistical aspects of large samples in large volumes
have been driving these new developments: from the simple constraint that measurements like
the number density in a luminosity function bin requires an accuracy of about 10\%,
with 10 such bins needed per redshift interval, in several redshift intervals to trace evolution,
for different types of galaxies, and different types of environments, one is quickly led to
define surveys with a requirement of more than $10^4-10^5$ galaxies with spectroscopic redshifts. 

Multi-object spectroscopy is undoubtedly the major technical development which has
enabled large redshift surveys. Until the late 1980s, spectroscopy was conducted object
by object, hence our view was biased towards a few rather extreme and not representative
samples (e.g. radio-galaxies, brightest cluster galaxies,...). 
Multi-fibre spectrographs have been developed to conduct large spectroscopic surveys of 
the nearby universe like the 2dFGRS (Colless et al. 2001), and the SDSS (York et al. 2000).
At higher redshifts, the first successful attempts
to use multi-slit masks on then distant galaxies happened in the late 80s early 90s with the Low Dispersion
Spectrograph at the AAT (Colless et al. 1990) and the MOS-SIS spectrograph at the CFHT (Le F\`evre et al. 1994).
The Canada France Redshift Survey (CFRS) used MOS-SIS to produce for the first time a 
sample of $\sim600$ galaxies with measured spectroscopic redshifts $0<z<1.2$ (Le F\`evre
et al., 1995), producing a landmark study with the discovery of the strong evolution
in luminosity and luminosity density (Lilly et al. 1995, Lilly et al. 1996, Madau et al. 1996). 

This impressive leap forward has in turn prompted a new generation of 
deeper spectroscopic redshift surveys, pushing the census of the galaxy  
population in larger volumes and going deeper beyond redshift $z\sim1$,
driving the development of multi-slit spectrographs 
on the new generation of 8--10m telescopes. On the Keck telescopes LRIS has
enabled the first large survey of Lyman-Break Galaxies (Steidel et al. 1996),
and has been followed by DEIMOS in 2002 used e.g. for the DEEP2 survey (Davis et al. 2003).
On Gemini-North, the GDDS (Abraham et al. 2004) has been enabled by GMOS (Hook et al. 2003). On the VLT, FORS-1
and FORS-2 upgraded with slit-masks have been used e.g. for the FDF (Noll et al. 2004) and GMASS (Cimatti et al.
2008). The FORS instruments have been followed on the VLT by 
the higher multiplex VIMOS, put in operations in 2002 (Le F\`evre et al., 2003), and which led to a number of deep redshift
surveys: the VVDS (Le F\`evre et al., 2005a), zCOSMOS (Lilly et al. 2007), 
a survey in the GOODS fields (Popesso et al. 2009), and the on-going
VIPERS (Guzzo et al. 2013), VUDS (Le F\`evre et al. in prep.), and zUDS (Almaini
et al. in prep.). 

These surveys each use different ways to pre-select targets and are, in many ways, complementary. 
Several selection methods have been used to define targets for deep spectroscopic surveys.
The easiest to implement is to select samples based on the luminosity of the sources
in a given bandpass, like magnitude-limited or line-flux limited surveys. 
This is used by a number of surveys selecting mainly in the I-band 
(CFRS, VVDS, zCOSMOS-wide,...) or in near-IR bands (GMASS, GDDS). 
When the emphasis is on a particular high redshift range, then additional
selection must be added to a magnitude limit, with colour-colour
selection used at $z\sim1$ (e.g. DEEP2, VIPERS), or, at higher redshifts,
using the BzK criterion (Daddi et al. 2004) to select populations with $1.4<z<2.5$ (e.g. zCOSMOS-deep)
or the Lyman-break technique to select populations at $z>2$
(Steidel et al. 1996, Steidel et al. 1999, ...). Another powerful
selection method is the use of narrow band imaging to preselect $Lyman-\alpha$
emitters (LAE) followed by multi-slit spectroscopy (e.g. Ouchi et al. 2010). 
While each method has its pros and cons, the key constraint for all
surveys is to maintain a tight
control of the selection function within the parameter space probed
by a survey, to be able to relate the observed population to the global
underlying population and derive absolute volume quantities like the
star formation rate density, the stellar mass density, the merger rate, etc.. 

Detailed sub-population studies fully depend on
the availability of large spectroscopic redshift surveys and their selection function,
providing fair and representative samples.
It is noteworthy that the significance of integral field spectroscopy studies of high redshift
galaxies at $z\sim>1$ using the H$\alpha$ line redshifted into the infrared
(e.g. Forster-Schreiber et al. 2009, Epinat et al. 2009, Law et al. 2009, Contini et al. 2012)
fully depends on the quality of the parent spectroscopic redshift survey.  
If one combines the requirements to investigate a galaxy population within a given star 
formation or stellar mass range, over a specific cosmic time, with redshifts 
such that H$\alpha$ and other important lines are away from the
sky OH emission lines, and with a sufficiently bright star to assist in adaptive optics
corrections, a survey with $10^4$ spectroscopic redshifts will provide only a few
hundred good targets for follow-up integral field spectroscopy.   

The VIMOS VLT Deep Survey (VVDS) has been conceived to provide a major contribution 
to deep spectroscopic galaxy redshift surveys.
It is based on a pure i-band magnitude selection, enabling a large range in redshift, and a large sample of
objects with spectroscopic redshifts for detailed statistical studies of the
high redshift galaxy and AGN population. The VVDS has produced a number of landmark studies at redshifts 
1 and above, based on a complete census of the galaxy population (e.g. Ilbert et al. 2005, Le F\`evre et al. 2005b,  
Cucciati et al. 2006, Arnouts et al. 2007, Franzetti et al. 2007, Pozzetti et al. 2007, Tresse et al. 2007, Guzzo et al. 2008, Zucca
et al. 2008,  de Ravel et al. 2009, Meneux et al. 2009,...). 
The richness of the VVDS sample is still producing new original analysis of the high
redshift population (e.g. Cassata et al. 2011, de la Torre et al. 2011, L\'opez-Sanjuan et al. 2011,
Cucciati et al. 2012, Cassata et al. 2013,...). 

In this paper we present the main properties
of the VVDS survey as now completed with 35\,016 galaxies and AGN with spectroscopic
redshifts covering a redshift range $0<z<6.7$, as a {\it Legacy} reference for existing and future analysis.
A summary of the VVDS is presented in Section \ref{overview}.
In section \ref{observations} we describe the observational methods
used to conduct the three increasingly deeper {\it wide}, {\it deep},
and {\it ultra-deep} i--band magnitude limited surveys down to
$i_{AB}=24.75$. 
We use independently observed samples to assess the reliability
of our redshift measurements.
We describe the VVDS surveys in Section \ref{sur}, and
the detailed selection function of the samples is computed in
Section \ref{compl}.
The global properties of the VVDS samples are presented in Section \ref{sample}, 
and we compare to other surveys in Section \ref{comparison}. 
We present the final VVDS data release in Section \ref{release}
and summarize the VVDS survey in Section \ref{summary}.
All magnitudes are given in the AB system unless specified,
and we use a standard Cosmology with $\Omega_M=0.3$,
$\Omega_{\Lambda}=0.7$ and $h=0.7$.


\section{Survey overview}
\label{overview}

The VIMOS VLT Deep Survey (VVDS) is a purely magnitude-limited spectroscopic redshift survey
with three complementary surveys: the VVDS-{\it Wide}, the VVDS-{\it Deep}, and
the VVDS-{\it Ultra-Deep}. The VVDS-Wide covers 8.7 deg$^2$ in five fields (0226-04, 1003+01, 1400+05, 2217+00, ECDFS), 
with 25\,805 spectroscopic galaxy redshifts down to $I_{AB}=22.5$. The VVDS-Deep has 
assembled 11\,486 galaxy redshifts in 0.74 deg$^2$ in the 0226-04 and ECDFS fields down to $I_{AB}=24$.
The VVDS-UltraDeep has pushed to $23 \leq i_{AB} \leq 24.75$, and obtained 938 galaxy 
redshifts in a 512 arcmin$^2$ area. Serendipitous discovery of Lyman-$\alpha$ emitters
in the slits of the VVDS Deep and Ultra-Deep i-band selected targets has added another 133 galaxies with $2<z<6.7$
(Cassata et al. 2011). 

\begin{figure*}
   \centering
   \includegraphics[width=15cm]{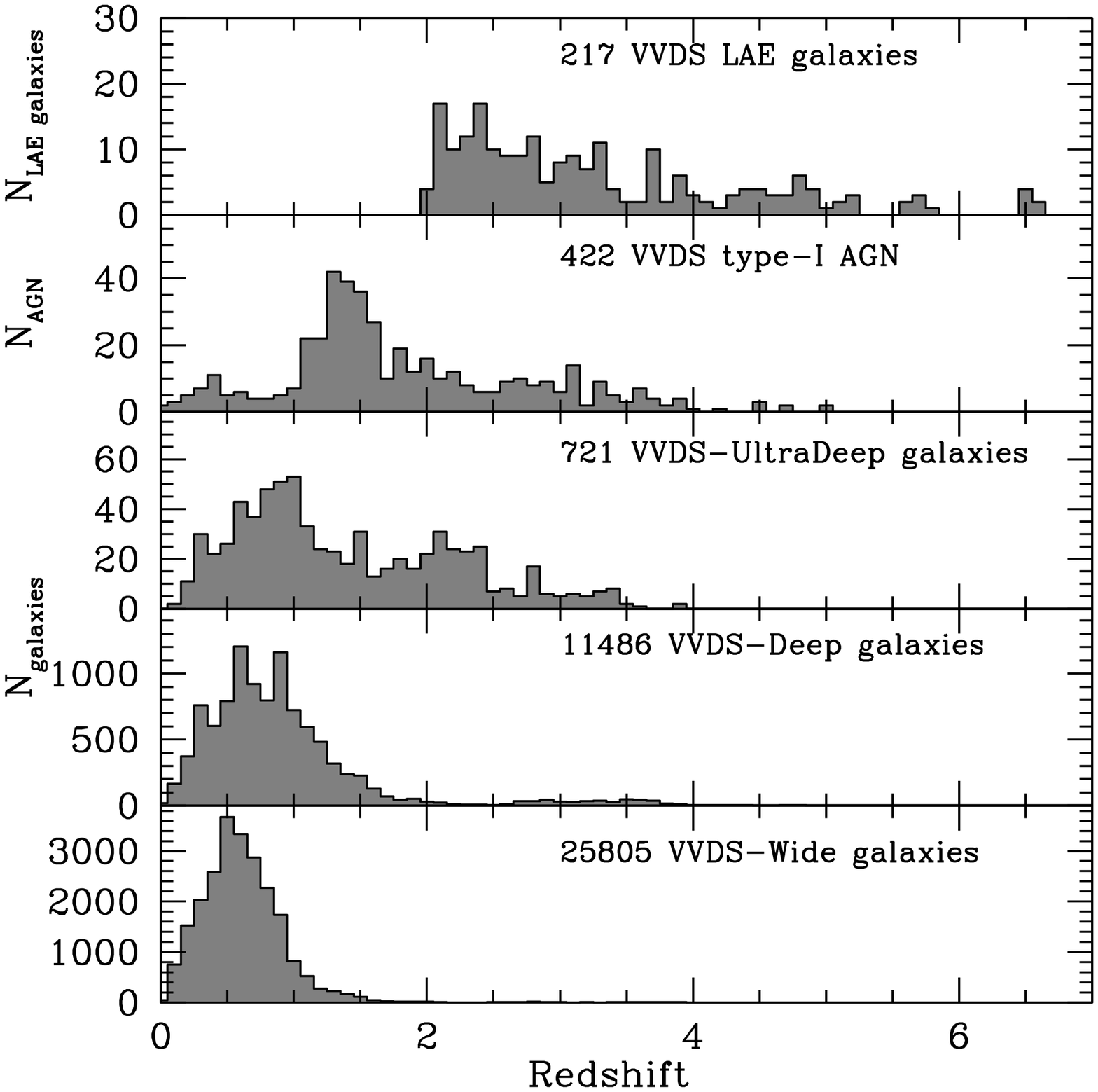}
      \caption{The redshift distributions from the 35\,016 galaxies in the VVDS surveys: 
               25\,805 galaxies in the VVDS-Wide (including 3\,768 from the VVDS-Deep satisfying the
               $17.5 \leq I_{AB} \leq 22.5$ selection), 11\,486 galaxies in the
               VVDS-Deep,  721 galaxies in the VVDS-UltraDeep (excluding those re-observed
               from the VVDS-Deep), and 217 Lyman-$\alpha$ emmitters (133 found
               serendipitously, the other galaxies coming from the Deep and Ultra-Deep samples).
               The redshift distribution of the 422 type-I AGN found in the VVDS surveys
               is shown on the top panel. 
              }
         \label{nz}
   \end{figure*}

In total the VVDS has obtained 34594 galaxy redshifts, 422 AGN type-I (QSO) redshifts,
and covers a large redshift range $0<z<6.7$. 
Such a large redshift coverage 
enables a detailed study of galaxy evolution over more than 13 billion years
of cosmic time, based on a simple sample selection.
The complete redshift distributions of the VVDS surveys are presented in Figure \ref{nz}.
The median redshifts for the Wide, Deep and Ultra-Deep surveys are
z=0.55, 0.92 and 1.38, respectively.
We summarize the total number of
measured spectroscopic redshifts for each VVDS survey in Table \ref{vvds}, 
and list the number of galaxies for several redshift ranges in Table \ref{znum}.
We emphasize that, although this domain is recognized to be difficult, the VVDS samples
successfully identify galaxies in the 'redshift desert' $\sim1.5<z<2.5$, allowing
detailed investigations of individual galaxies at this epoch 
(it has enabled e.g. the MASSIV 3D spectroscopy survey in $1<z<2$, Contini et al. 2012).  

As described in Section \ref{irselect},  magnitude-selected samples at  wavelengths other than i-band
can be easily extracted from the VVDS surveys. The VVDS-Deep in the 0224-04 field provides  
nearly magnitude-complete samples of 7830, 6973, and 6172 galaxies with redshifts down to $J_{AB}=23$, $H_{AB}=22.5$,
and $Ks_{AB}=22$, respectively.
Flux-limited samples can be extracted using multi-wavelength data,
radio (VLA, down to $17\mu Jy$ at 1.4 Ghz), X-rays (XMM, $\sim 5\times 10^{-15}~{\mathrm {erg~cm^{-2}~s^{-1}}}$ in the 0.5 2 keV band), 
mid-IR (Spitzer, in 3.6, 4.5, 5.6, 8 and 24 $\mu$m, down to $3.7\mu Jy$ at 3.6$\mu$m), UV (Galex, FUV and NUV, down to $NUV_{AB}=24.5$)
or far-IR (Herschel, at 250, 350 and 500 $\mu$m, down to $\sim 13 mJy$),
together with the VVDS magnitude selected surveys, as described in Section \ref{ima}.

   \begin{table*}
      \caption[]{VVDS samples}
      \[
        \begin{array}{lccrrrrl}
           \hline \hline
            \noalign{\smallskip}
Sample       &  $Selection$ & \bar{z}   & $All galaxies$   & $Galaxies with$              & $Galaxies with$       & $AGN$ & $References$ \\
             &              & [z_{min},z_{max}] & $with redshifts$ & $reliable redshifts$         & $tentative redshifts$ &       &  \\
             &              &           &                  & $(flags 1.5$^a$,2,3,4,9)$    & $(flag 1)$            &       &  \\
            \noalign{\smallskip}
            \hline
            \noalign{\smallskip}
$Wide$^b       &  17.5 \leq I_{AB} \leq 22.5 & 0.55 \smallskip[0.05,2] &  22\,037 & 17\,150 & 4\,887   & 304 & $Garilli et al. (2008); this paper$ \\ 
$Deep-Wide$^c &  17.5 \leq I_{AB} \leq 22.5  &               &   3\,768 &  3\,535 &    233   &  69 & $Le F\`evre et al. (2005a); this paper$\\
            \noalign{\smallskip}
            \hline
            \noalign{\smallskip}
$Total Wide$  & 17.5 \leq I_{AB} \leq 22.5 &                 & 25\,805 & 20\,685 & 5\,120   &  373 & \\
            \noalign{\smallskip}
            \hline
            \noalign{\smallskip}
$Deep$^d       &  17.5 \leq I_{AB} \leq 24  &  0.92 \smallskip[0.05,5] & 11\,486 &  9\,641 & 1\,845   & 115 & $Le F\`evre et al. (2005a); this paper$\\
$Ultra-Deep$ & 23 \leq i_{AB} \leq 24.75   &  1.38 \smallskip[0.05,4.5] &  938 &   756 &  182   & 3   & $this paper$ \\
$Lyman-$\alpha $~emitters$^e & >1.5 \times 10^{−18} erg/s/cm^2  & 3.6 \smallskip[2,6.7]  & 133 &   133 &    -   & -   & $Cassata et al. (2011)$ \\
Lyman-\alpha ~emitters^f   &                                  &  \it{3.3 \smallskip[2,6.7]} & \it{217} &  \it{217} &    -   & -   &  \\
            \noalign{\smallskip}
            \hline
            \noalign{\smallskip}
$TOTAL all VVDS $^g    & - & - & 34\,594 & 27\,680 & 6\,914   & 422 & \\
            \noalign{\smallskip}
            \hline
         \end{array}
      \]
\begin{list}{}{}
\item[$^{\mathrm{a}}$] For Ultra-Deep only
\item[$^{\mathrm{b}}$] In fields 1003+01, 1400+05 and 2217+00
\item[$^{\mathrm{c}}$] Limiting the VVDS-Deep to $I_{AB}\leq22.5$ in the 0226-04 and ECDFS fields
\item[$^{\mathrm{d}}$] In fields 0226-04 and ECDFS; includes the "Deep-Wide" sample limited to $I_{AB}\leq22.5$
\item[$^{\mathrm{e}}$] Serendipitous LAE emitters only
\item[$^{\mathrm{f}}$] Serendipitous LAE and LAE from Deep and Ultra-Deep 
\item[$^{\mathrm{g}}$] Summing the Wide, Deep, Ultra-Deep, and LAE surveys 
\end{list}
\label{vvds}
   \end{table*}

   \begin{table*}
\begin{center}
      \caption[]{Numbers of galaxies with spectroscopic redshifts in the VVDS surveys}
      \[
        \begin{array}{lrrrrrr}
           \hline \hline
            \noalign{\smallskip}
            Survey    &  0< z_{spec} \leq 0.5 & 0.5<z_{spec}\leq 1  & 1<z_{spec} \leq 1.5 & 1.5<z_{spec} \leq 2  & 2 < z_{spec} \leq 3  & z_{spec}>3 \\
                        &  $all / reliable$       & $all / reliable$      & $all / reliable$      & $all / reliable$       & $all / reliable$       & $all / reliable$ \\
            \noalign{\smallskip}
            \hline
            \noalign{\smallskip} 
            $Wide$^b      &  8\,126 / 6\,619          & 12\,175 / 9\,598        & 1\,476 / 891           & 162 / 26             & 62 / 13           & 36 / 3 \\

            $Deep$^c      &  2\,510 / 2\,189          & 5\,742 / 5\,169         & 2\,364 / 1\,941          & 351 / 138            & 211 / 70            & 308 / 134 \\
            $Ultra-Deep$ &   107 / 90            &   236 / 202         &   184 / 152         & 132 / 98              & 238 / 181            & 41 / 33 \\
            $Ly$\alpha $~emitters$^d &   0         &     0               &     0               &  0                   & 50 / 50             & 83 / 83 \\
            \noalign{\smallskip}
            \hline
            \noalign{\smallskip}
	    $TOTAL$       &   10\,743 / 8\,898        &   18\,153 / 14\,969     & 4\,024 / 2\,984         &  645 / 262           & 561 / 314           & 468 / 253 \\
            \noalign{\smallskip}
            \hline
         \end{array}
      \]
\begin{list}{}{}
\item[$^{\mathrm{a}}$] Serendipitous LAE emitters only 
\item[$^{\mathrm{b}}$] In fields 1003+01, 1400+05 and 2217+00
\item[$^{\mathrm{c}}$] In fields 0226-04 and ECDFS
\end{list}
\label{znum}
\end{center}
   \end{table*}

\section{Observations and redshift measurement reliability}
\label{observations}

\subsection{Survey fields and area}

The survey fields have been chosen to be on the celestial equator to enable
visibility of any two field at any time of the year, at locations with low galactic extinction.
The fields positions and the survey modes applied on each are listed in Table \ref{fields}.

The 0226-04 field is the most observed field in the VVDS hosting most of the 
Deep and all of the Ultra-Deep surveys. It is included in the CFHTLS (Cuillandre et al., 2012)
and XMM-LSS (Pierre et al., 2004) imaging areas, and other multi-wavelength data are available as described
in Section \ref{ima}.
This field was defined and observed before the Subaru
SXDF field which has a field centre only 2 degrees away at 0218-05, leaving 
only one degree between them, such that joining these two fields 
with bridging photometry and spectroscopy would offer the possibility for a unique large extragalactic field
covering more than 3 deg$^2$. This 0226-04 field was to become the COSMOS field (Scoville et al. 2007),
but this did not happen at the request of the STScI and ESO directors to limit
R.A. overload on these facilities, already committed to the GOODS areas including
the ECDFS at R.A.=3h32m. The VVDS-10h field was instead selected as the COSMOS
field to be observed with HST, despite having at that time limited multi-wavelength data
and multi-object spectroscopy available. 

The 2217+00 field is in Selected Area SA22, a well studied area revisited by several
spectroscopic surveys (e.g. Lilly et al. 1991, Lilly et al. 1995, Steidel et al. 1998), 
and the VVDS area is now included in the CFHTLS wide imaging survey, with extended
redshifts from the VIPERS survey (Guzzo et al. 2013).
 
The VVDS-Wide field at 1003+01 has evolved into the COSMOS field (Scoville et al. 2007),
which was slightly displaced to 1000+02 to avoid some large galactic extinction areas
found when more accurate extinction maps were made available after the initial VVDS
field selection and imaging survey. The VVDS field is then
partially overlapping and extending the area covered with spectroscopy by the zCOSMOS-Wide
survey (Lilly et al. 2007).

The 1400+05 field is a high galactic latitude field, and the ECDFS was added 
as a reference field with VVDS redshifts made rapidly public to a
broad community (Le F\`evre et al. 2004a).

The fields location and covered area for each are indicated in Table \ref{fields}.


   \begin{table*}
\begin{center}
      \caption[]{VVDS fields and exposure times}
      \[
        \begin{array}{lllcccccccc}
           \hline \hline
            \noalign{\smallskip}
            Field      &  \alpha_{2000} & \delta_{2000} & b  & l  & $Survey mode$ & $Area$ & $VIMOS setup$ & T_{exp} (h) \\
            \noalign{\smallskip}
            \hline
            \noalign{\smallskip}
            1003+01 &  10h03m00.0s & +01\deg30\arcmin00\arcsec & 42.6 & 237.8 & Wide & 1.9 $deg$^2 & LRRED & 0.75 \\
            1400+05 &  14h00m00.0s & +05\deg00\arcmin00\arcsec & 62.5 & 342.4 & Wide & 2.2 $deg$^2 & LRRED & 0.75 \\
            2217+00 &  22h17m50.4s & +00\deg24\arcmin00\arcsec & -44.0 & 63.3 & Wide & 4.0 $deg$^2 & LRRED & 0.75 \\
            \noalign{\smallskip}
            \hline
            \noalign{\smallskip}
            0226-04 $~Deep$ &  02h26m00.0s & -04\deg30\arcmin00\arcsec & -58.0 & -172.0 & Deep & 0.61 $deg$^2 & LRRED & 4.5 \\
            $ECDFS$ &  03h32m28.0s & -27\deg48\arcmin30\arcsec & -54.5 & 223.5 & Deep & 0.13 $deg$^2           & LRRED & 4.5 \\
            \noalign{\smallskip}
            \hline
            \noalign{\smallskip}
            0226-04 $~Ultra-Deep$ & 02h26m28.8s & -04\deg23\arcmin06\arcsec  & 58.0 & -172.0  & Ultra-Deep  & 512 $arcmin$^2 & LRBLUE & 18 \\
                                  &             &                            &      &         &             &                & LRRED  & 18 \\
            \noalign{\smallskip}
            \hline
         \end{array}
      \]
\label{fields}
\end{center}
   \end{table*}

\subsection{VIMOS on the VLT}
\label{vimos}

The VIsible Multi-Object Spectrograph (VIMOS) is installed on the European Southern Observatory
Very Large Telescope unit 3 Melipal. It was commissioned in 2002 (Le F\`evre et al. 2003), and has been 
in regular operations as a general user instrument since then. VIMOS is a wide field imaging multi-slit spectrograph, hence offering
broad band imaging capabilities in u,g,r,i, and z bands, as well as multi-slit spectroscopy with spectral
resolution ranging from $R\simeq230$ to $R\simeq2500$ (1 arc-second slits), and spectral coverage in
$3\,600 \leq \lambda \leq 10\,000$\AA ~depending on the dispersing element used (grism / VPH). VIMOS also offers a wide field
integral field spectroscopy capability in a field ranging from $27\times27$arcsec$^2$ to $54\times54$arcsec$^2$.
The global VIMOS throughput at 6\,000\AA ~without the detector and dispersing element is an excellent 78\%. 

\begin{figure}
   \centering
   \includegraphics[width=8cm]{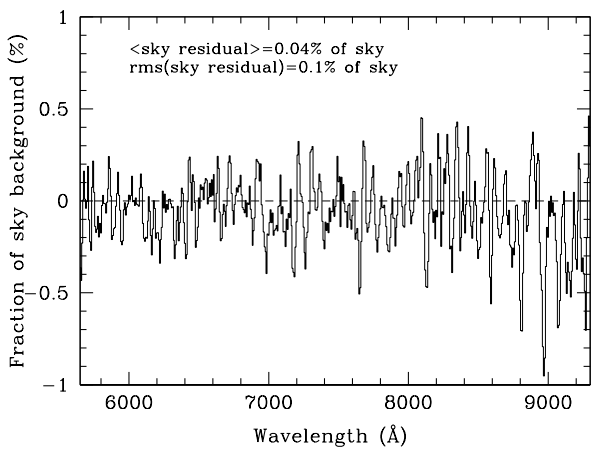}
      \caption{Sky background subtraction accuracy expressed as the ratio of the sky subtracted spectrum
               over the observed sky spectrum, obtained from the 18h stack of VIMOS-LRRED observations
               for the VVDS-UltraDeep survey. 
              }
         \label{sub}
   \end{figure}

The multi-slit spectroscopy capability has been specifically designed to offer a large multiplex, with the
number of individual slits/objects observed simultaneously ranging from $\sim200$ to $\sim800$ at 
high to low spectral resolution, respectively. Multi-slit masks are cut in Invar sheets to excellent accuracy
(a few microns, less than one hundredth of an arc-second) using a dedicated laser machine (Conti et al. 2001). 
Besides the global spectrograph throughput, the ability to observe faint targets relies heavily on the sky signal subtraction accuracy.
From our multi-slit observations we consistently reach a sky background subtraction accuracy 
of better than $\sigma_{sky residual}\simeq0.1$\% of the sky background intensity, 
even when stacking up to 18h of observations, as shown in Figure \ref{sub}.

The VVDS Wide and Deep surveys have been conducted with the LRRED grism covering $5500 \leq \lambda \leq 9350$\AA, while the
VVDS Ultra-Deep survey has used the LRBLUE and LRRED grisms to cover $3650 \leq \lambda \leq 9350$\AA, both with
slits one arc-second wide. This provides a spectral resolution $R=230$. At this resolution, the detectors can accommodate
3--4 full length spectra along the dispersion direction, and, given the projected space density of VVDS targets, 
more than 400 object-slits per pointing have been observed for the Wide survey, and more than 500 object-slits for the DEEP 
and Ultra-Deep surveys. More details can be found in Le F\`evre et al. (2005a), and Garilli et al. (2008). 



\subsection{Photometric i--band selection: imaging}

All the VVDS spectroscopic sample is I--band (Wide and Deep) or i--band  (Ultra-Deep) selected. 
In support of the VVDS an imaging campaign has been conducted at the CFHT using the CFH12K camera,
in BVRI bands (Le F\`evre et al. 2004b). The main requirements for this imaging survey was to be 
deep enough to select targets down to $I_{AB}=22.5$ and to cover a total of 16 deg$^2$
for the Wide survey, and reaching down to $I_{AB}=24$ and cover 1 deg$^2$ for the Deep survey.
Integration times in the I--band of one hour on the Wide survey and 3h on the Deep survey
led to limiting magnitudes of $I_{AB}=24.8$ and $I_{AB}=25.3$ at $5\sigma$ in a 3 arc-second aperture
(Le F\`evre et al. 2004b). As described in McCracken et al. (2003), the depth of the imaging survey
ensures a 100\% completeness in detecting objects down to the limiting magnitude of
the spectroscopic survey. 

For the Ultra-Deep survey, the release no.3 of the CFHT Legacy Survey i--band imaging on the 0226-04
field has been used to select targets down to $i_{AB}=24.75$. The Megacam i--band filter has a bandpass
close to but slightly different from the CFH12K I--band used for the Deep and Wide surveys, warranting
a specific notation throughout this paper. The imaging depth 
of the CFHTLS reaches  $i_{AB}=25.5$ at $5\sigma$ in a 3 arc-second aperture, sufficiently
deep to avoid any imposed imaging selection bias.

The I--band or i--band selections are based on SExtractor (Bertin and Arnouts, 1996)
$mag-auto$ a Kron-magnitude approximating a total magnitude. Two-band colors are computed in 3 arcsecond apertures
to ensure that the spectral energy distribution of each galaxy is measured in the same physical size in
each band, as well as to maximize flux and to minimize the contamination correction residuals from nearest neighbors.
Besides the i--band used for target selection, multi-band imaging and multi-wavelength data have 
been assembled in the VVDS survey fields as described in Section \ref{ima}.

\subsection{Spectroscopic data processing, redshift measurement, and reliability flag}
\label{zmeas}

The multi-slit spectroscopy data processing has been performed using the VIPGI data processing environment
(Scodeggio et al., 2005). It consists of 2D spectra extraction, sky-subtraction, combination
of individual 2D spectra, 1D spectra tracing and extraction, wavelength and flux calibration
of the 2D and 1D spectra. 
The redshift measurements have been performed in several steps
using the EZ engine developed within the VVDS (Garilli et al., 2010), based on cross-correlation
with spectra templates and augmented by knowledge-based software.

The redshifts are measured separately by 2 persons in the team, and reconciled
from a face to face confrontation.
This confrontation  is designed to smooth-out the possible biases of individual observers
and produce an homogeneous reliability assessment by means of a flag. 
Each redshift measurement has a spectroscopic flag associated to it, indicating the probability 
for this particular redshift to be right. This method was originally pioneered by the 
CFRS (Le F\`evre et al. 1995), and since then has been used by other major surveys 
besides the VVDS like zCOSMOS
(Lilly et al. 2007) or VIPERS (Guzzo et al. 2013). This probabilistic approach
has been proven over these large surveys to be both reliable and easy to
handle in terms of evaluating the selection function of the survey 
with various sampling rates as described in Section \ref{selra}.

The flag may take the following
values:
\begin{itemize}
\item 4: 100\% probability to be correct
\item 3:  95--100\% probability to be correct
\item 2:  75--85\% probability to be correct
\item 1:  50--75\% probability to be correct
\item 0:  No redshift could be assigned
\item 9:  spectrum with a single emission line. The redshift given is the most probable
          given the observed continuum, it has a $\sim80$\% probability to be correct. 
\end{itemize}
These flag probabilities are examined in more details below.

From this basic flag list, more specific flags have been build using a second digit
in front of the reliability digit.
The first digit can be "1" indicating that at least one emission
line is broad, i.e. resolved at the observed spectral resolution, or "2" if 
the object is not the primary target in the slit but happens to fall in the slit
of a primary target by chance projection, and hence provides a spectrum.
For the VVDS-UltraDeep, we have added a flag 1.5 corresponding to objects for 
which the spectroscopic flag is "1", and the 
spectroscopic and photometric redshifts match to within $dz=0.05\times(1+z)$.

This statistical method has been extensively
tested from independent repeated observations of several hundred  objects, consistently
providing similar statistical reliability estimates for the different flag categories 
(Le F\`evre et al. 1995, Le F\`evre et al. 2005a, Lilly et al. 2009, Guzzo et al. 2013). 

We have consolidated this redshift probability scheme from several lines of evidence.
The duplicate observations obtained in the VVDS-Deep have been presented in Le F\`evre et al. (2005a).
About 386 objects have been observed twice in the ECDFS and 0226-04 fields, processed and redshifts measured independently.
More recently, 558 objects from the VVDS-Wide have been re-observed with VIMOS in the context of the
VIPERS survey (Guzzo et al. 2013), and 88 VVDS galaxies have been observed in 
the near-IR with VLT-SINFONI for the purpose of the MASSIV survey (Contini et al. 2012),
providing fully independent redshift measurements.

\begin{figure*}
   \centering
   \includegraphics[width=15cm]{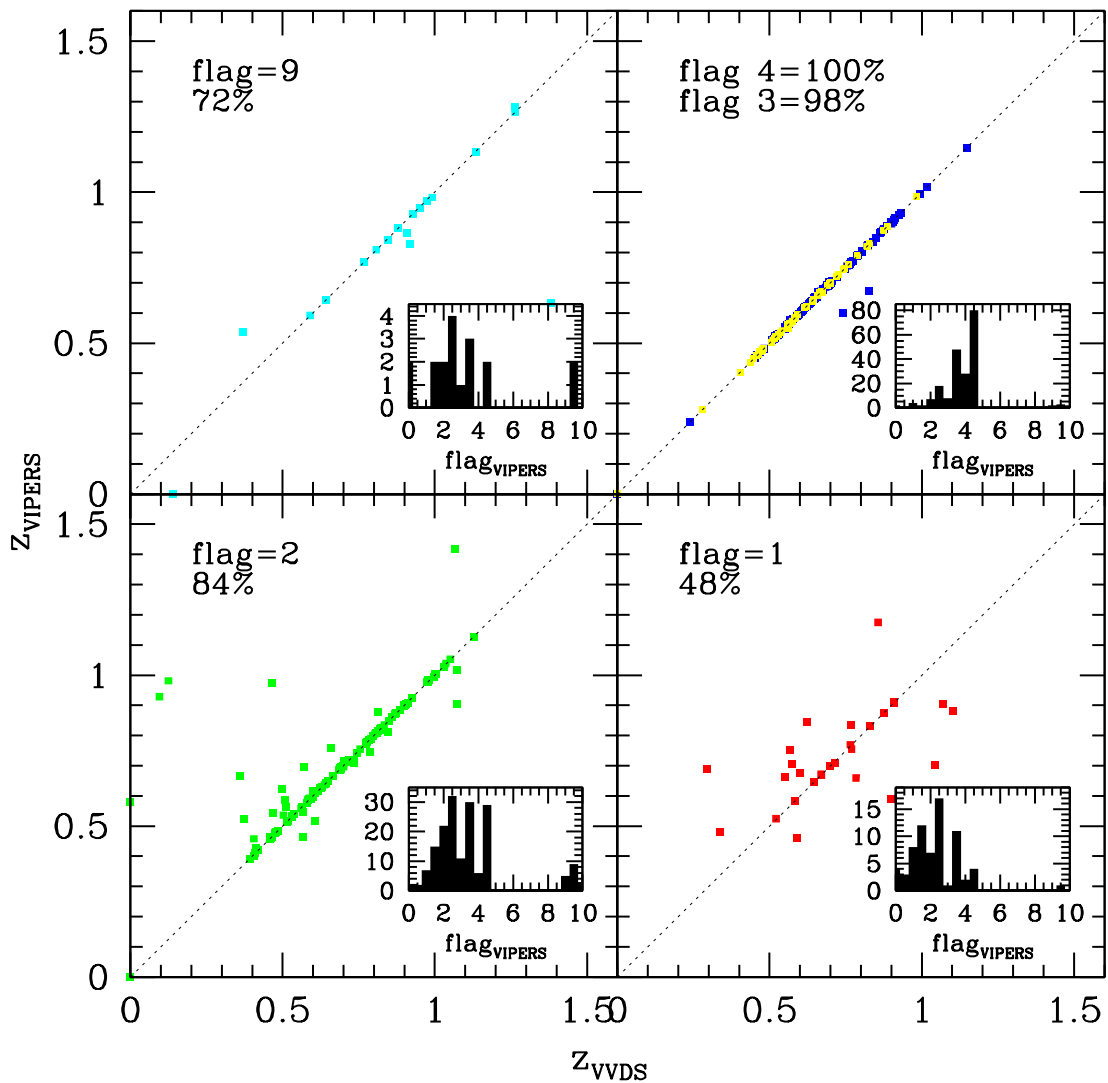}
      \caption{Comparison of independent redshift measurements and classification of 558 galaxies 
               in VVDS-Wide and in the VIPERS survey (Guzzo et al., 2013). The comparison
               is made for each of the 4 categories of spectra in the VVDS: flags 1, 2, 3 and 4, and 9.
               The probability for the redshift of each class of galaxy is computed as the 
               fraction of galaxies which agree between VVDS and VIPERS to $|dz| \leq 0.0025\times(1+z)$, and is indicated in the upper
               left corner of each panel, using as a reference the most reliable redshift flags
               from the VIPERS survey (the small histogram inset in each panel shows the VIPERS flag distribution).  
	       Using only the most reliable flags, the sigma of the redshift difference between 
               VVDS and VIPERS redshift measurements normalized to the expansion factor (1+z)
               is $\sigma_{dz}=0.00072$ which translates to an individual redshift measurement uncertainty of $\sigma_v=202$km/s.
              }
         \label{reobs_vipers}
   \end{figure*}

The VIPERS survey uses VIMOS with the red, higher QE, CCDs installed in 2010  with a similar exposure time
to the VVDS, hence providing improved S/N at fixed exposure time. 
We have compared the 558 VVDS redshift measurements in common to VIPERS in Figure \ref{reobs_vipers}. 
For each flag category of galaxies in the VVDS, we compare the VVDS redshift measurements  to
those of VIPERS, considering that they agree if the velocity difference is $|z_{VVDS}-z_{VIPERS}| \leq 0.0025\times(1+z)$,
consistent at $\sim3\sigma$ with the accuracy in redshift measurement (see below).
As VIPERS observations have better S/N than the VVDS we have used the VIPERS galaxies with 
flags 2, 3, 4 as the exact redshift reference as they have a probability to be right from 94.8 to 100\% (Guzzo
et al. 2013), and we have defined the probability for a 
VVDS redshift to be correct as the ratio of galaxies with a redshift agreement between VVDS and these VIPERS galaxies
over the total number of galaxies with redshifts for each VVDS flag category.
We find the following probabilities that the VVDS redshift measurements are correct:
flag 1: 48\%, flag 2: 84\%, flag 3: 98\%, flag 4: 100\%, and flag 9: 72\%.

\begin{figure}
   \centering
   \includegraphics[width=8cm]{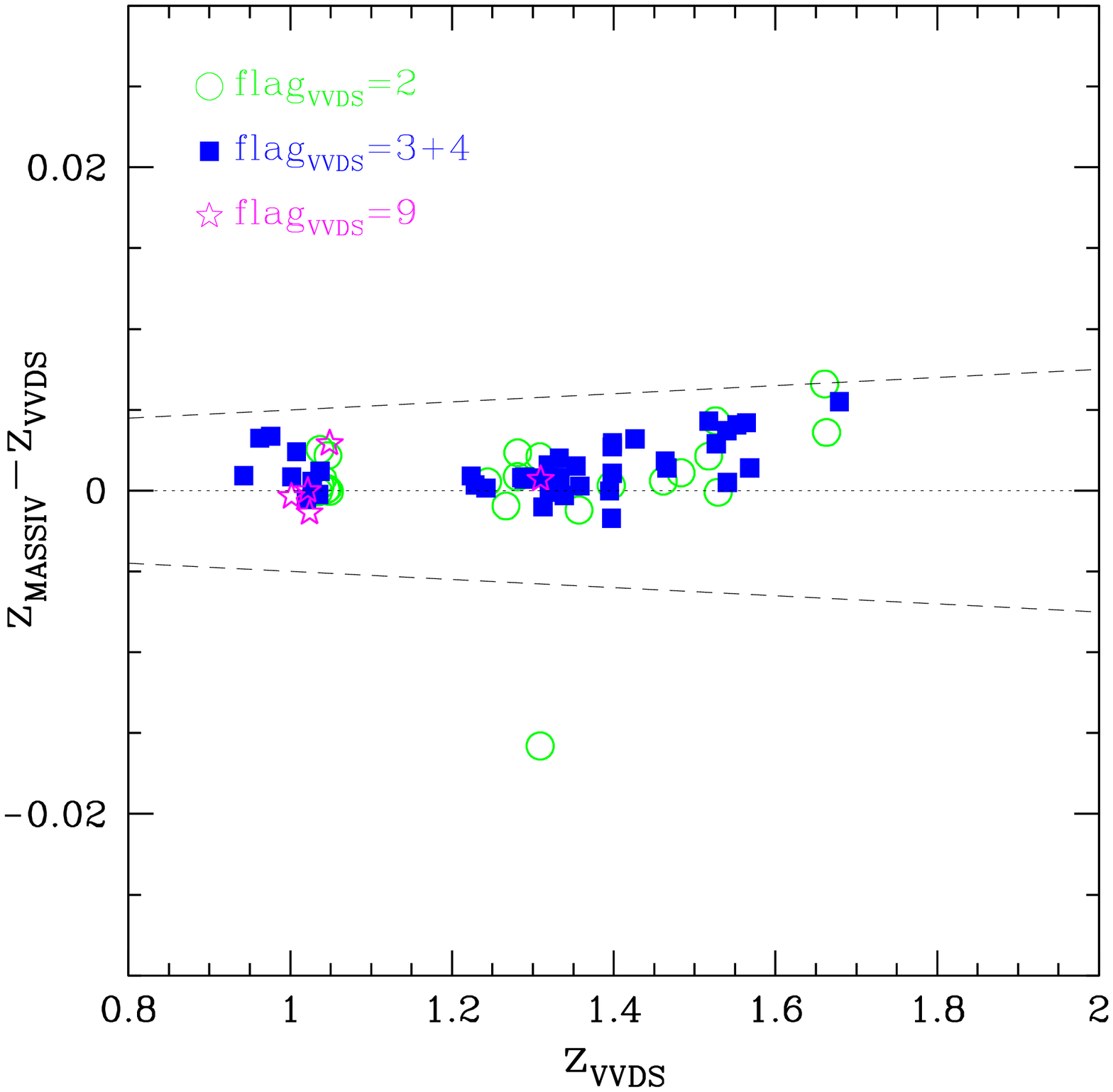}
      \caption{Redshift difference $dz=z_{MASSIV}-z_{VVDS}$ between the independent redshift measurements of the VVDS
               and MASSIV surveys for 88 galaxies 
               in the VVDS and in the MASSIV survey (Contini et al., 2012). Flags 2 are represented
               with circles, flags 3 and 4 with squares, and flag 9 with stars.
               Dashed lines identify $dz=0.0025\times(1+z)$. The sigma of the redshift difference between
               VVDS and MASSIV redshift measurements normalized to the expansion factor (1+z)
               is $\sigma_{dz}=0.00072$ which translates to an individual redshift measurement uncertainty of $\sigma_v=153$km/s.  
              }
         \label{reobs_mass}
   \end{figure}

The MASSIV survey is a targeted survey to study the kinematic properties of galaxies
with $1<z<1.8$ using integral field spectroscopy on the H$\alpha$ line in the J and H bands (Contini et al. 2012).
On the 88 galaxies selected from the VVDS, 30, 40, 13, and 5 have flags 2, 3, 4, and 9, respectively.
A total of 72 objects have H$\alpha$ or [OIII]5007\AA ~detected at a flux above $2 \times 10^{-17}$ erg/s/cm$^2$, 
16 are not, or only marginally, detected. All of the detected objects but one have the same
redshift as measured from the H$\alpha$ or [OIII]5007\AA ~in MASSIV and in the VVDS (Figure \ref{reobs_mass}).
For the 16 non detected objects, 2 have a flag 4 among a total of 12 flag 4 observed,  6 have a flag 3 over 40,
and 8 have a flag 2 over 30 observed. 
With the flag 3 and 4 being 98\% and 100\% correct, we infer that  6 over 52 objects with such flags 
have not been detected in the SINFONI spectra mainly because the H$\alpha$ or [OIII]5007\AA 
~fluxes are below the 4$\sigma$ limit $F(line) \leq 1 \times 10^{-17}$ erg/s/cm$^2$ (this value is somewhat
wavelength dependent because of the varing sky and instrument background). 
Making the reasonable hypothesis that 15\% of the flag 2 objects in MASSIV are undetected for the
same reason, we may then deduce that the success rate for redshifts with flag 2 is 
$(22/30+0.15)$ or 88\%. 
All the 5 galaxies with flag 9 from the VVDS have their redshift confirmed from MASSIV.
Combining the 23 VVDS flag 9 galaxies re-observed with MASSIV and VIPERS, we find that 19 have the correct
redshift, hence of probability of 83\%.
We add that another 3 VVDS galaxies at $z\sim3.5$ have been observed
with SINFONI in a pilot program for MASSIV (Lemoine-Busserolle et al. 2010),
including 2 flag 3 and one flag 4 object, their redshifts were all confirmed
from at least [OIII]5007\AA ~detection.

From the VVDS objects re-observed by VIPERS we derive the velocity difference distribution.
The 1-sigma of the distribution normalized to the expansion factor $(1+z)$ is $\sigma_{1+2}=0.00095$, 
or  $\sigma_{1+2}\sim285$ km/s between two objects, corresponding to a velocity error per object of 
$\sigma_v=\sigma_{1+2} / \sqrt{2}=202$km/s, while
for VVDS objects re-observed by MASSIV, we find $\sigma_{1+2}=0.00072$ hence $\sigma_v\sim153$km/s.
We find that between VVDS and MASSIV, using fully independent instrument setups,
the absolute velocity zero point differs by only $\Delta z=0.0003$. 
This corresponds to $\sim 40$ km/s at the mean
redshift of MASSIV, well within the expected difference given that the 
measurements are coming from two different instrument systems with different spectral
resolutions.

Furthermore, we have re-observed in the Ultra-Deep $\sim250$ galaxies with 
flags 0,1,2 from the Deep survey. After 18h integration, most of these flags have become flags 3 or 4.
The comparison of redshifts and flags from both the Deep
observations and the much deeper Ultra-Deep observations on these galaxies is
in full support of the probabilities listed above for each flag, as described in 
Section \ref{reobs}.

These measurements of the redshift flag probability are fully in line
with our earlier estimates (Le F\`evre et al. 2005a), as well as that observed
in other surveys with similar redshift measuring scheme (Le F\`evre et al. 1995, Lilly et al. 2007, 
Guzzo et al. 2013). 
This scheme enables a full
use of all the galaxies with a redshift and flag measurement, provided appropriate
care is given to weight galaxies with different flags when computing
volume quantities in relation to the observed parent population, as we discuss in
details in Section \ref{compl}. 

Summarizing this section, the VVDS is using a well characterized statistical redshift reliability estimator,
which enables a robust statistical treatment of the complete galaxy population with measured redshifts. 
We demonstrate that the flag 2, 3, 4 and 9 are highly reliable at a level from $\sim83$\% to $100$\%.
We consider that these objects are forming the best VVDS sample, and we have used it in
all the VVDS science analysis. In addition, as the selection function is well defined, objects with flag 1 
can also be used despite their 50/50 failure rate. We are therefore
quoting the total number of measured redshifts in the VVDS as all objects with a redshift
measurement, i.e. with a flag root 1, 2, 3, 4 and 9.
The statistical properties of this flag system serve as a basis to the weights
defined in Section \ref{selra}, used to derive volume quantities and their associated errors. 
We invite people using our data release or quoting numbers of observed galaxies 
with redshifts to take into consideration this powerful statistical redshift measurement treatment.

\subsection{Redshift reliability vs. quality}

This flag system is sometimes misinterpreted in the literature 
as an indication of spectra quality (e.g. the recent  surveys comparison 
by Newman et al. 2012, albeit with incorrect numbers regarding the VVDS), 
but this association is not appropriate.
Defining the quality of a single redshift measurement is not as straightforward as it may
seem. Several estimators each give a different flavour of 'quality', like
the S/N on the continuum, the number and S/N of emission lines, the
strength of the cross-correlation signal. All of these could only be exactly
quantified in the presence of constant noise properties as a function of wavelength,
but with the highly non-linear background subtraction process, all
of these indicators are biased in one way or another. While spectra with 
high continuum S/N, many spectral lines, a strong correlation signal, lead to a  
redshift measurement obvious to all observers, faint galaxy surveys going to the limit
have to deal with a mixture of these indicators, not always with 
the best 'quality' for all indicators. One could get galaxies with a low S/N on the
continuum but an obvious set of emission lines matching a single redshift;
an emission line may fall on a sky line and be missing while 
the correlation signal on the continuum is strong; or, importantly, the S/N on the continuum could be low
but one could have a strong correlation signal because a number of features
are only each weakly detected but add to support the correlation.   
As a result, 'quality' assessment is then often subjective, 
strongly correlated to the individual who made the measurement, and
difficult to compare from one survey to another. 
To the contrary, the probabilistic approach used in the VVDS
guarantees an homogeneous treatment of the redshift measurements.  

Several different notation schemes or quality estimates have been used in the literature.
We contend that they are not of the same nature, and that care must be taken when comparing them. 
From our VIMOS experience dealing with $>1.5 \times 10^5$ spectroscopic redshift measurements from 
the VVDS, zCOSMOS, and VIPERS surveys, it appears that it is illusory to aim at a classification 
of galaxy redshifts measurements into a scheme as basic as {\it good} and {\it bad}, but rather
that using a more continuous distribution of redshift reliability as described above
is more appropriate to this type of dataset.

\section{VVDS surveys}
\label{sur}

\subsection{VVDS-Wide}

The VVDS-wide has been observing targets selected from their apparent I-band magnitude $17.5 \leq I_{AB} \leq 22.5$.
A total of 24, 28 and 51 VIMOS pointings have been observed in three fields 1003+01, 1400+05 and 2217+00, respectively. 
Integration times of 45 minutes have been obtained with the LRRED grism. 
These observations have been extensively described in Garilli et al. (2008).
The final data release presented here adds the observations of 24 more pointings ($\sim8000$ galaxies) to the Garilli et al. (2008) data.
The full VVDS-Wide sample consists of all objects in the three fields 1003+01, 1400+05 and 2217+00, as well as
the objects with $17.5 \leq I_{AB} \leq 22.5$ in the 0226-04 and ECDFS fields, for a total of $25\,805$ galaxies
and $305$ type-I AGN. The total area covered is 8.7 square degrees, the redshift range $0.05 \leq z_{Wide} \leq 2$
for a mean redshift $\bar{z}=0.55$ (Le F\`evre et al., 2013) and a volume $\sim2\times10^7$h$^{-3}$Mpc$^3$. 
As, deliberately, no attempt was made to apply star-galaxy separation algorithms
prior to select spectroscopic targets, the VVDS-Wide contains a rather large fraction of 34\% of Galactic stars, mainly
in the lower Galactic latitude 2217+00 field.

\subsection{VVDS-Deep}
\label{deep}

The VVDS-Deep sample is based solely on I--band selection $17.5 \leq I_{AB} \leq 24$.
Total integration times of 4.5h have been obtained with the LRRED grism.
The VVDS-Deep observations of the 0226-04 field have been extensively described in Le F\`evre et al. (2005a)
and the ECDFS in Le F\`evre et al. (2004a).

\begin{figure}
   \centering
   \includegraphics[width=8cm]{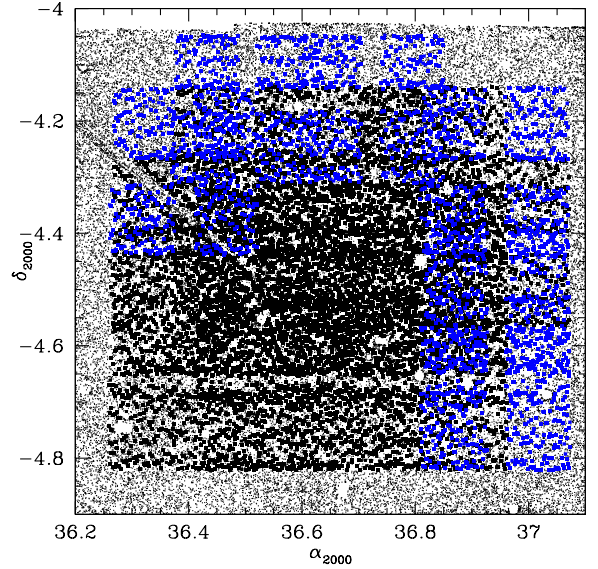}
      \caption{The distribution in ($\alpha, \delta$) of all galaxies observed in the 
               Deep $17.5 \leq i_{AB} \leq 24$ sample. Epoch 1 observations are indicated in black heavy symbols,
               observations from 8 additional VIMOS
               pointings are indicated in blue, and the
               photometric parent sample as dots.
              }
         \label{dist_deep}
   \end{figure}

The VVDS-Deep dataset now includes a sample of $\sim4000$ additional objects from 'epoch 2' 
observations of 8  VIMOS pointings. 
This sample has been observed and the 
data processed following the exact same procedure as described in Le F\`evre et al. (2005a). 
The full field observed from the VVDS-Deep
'Epoch 1' (Le F\`evre et al. 2005a) and 'Epoch 2' observations covers 2200 arcmin$^2$ as shown
in Figure \ref{dist_deep}.

This brings the total sample of objects observed in the VVDS-Deep
to 12\,514: 11\,486 are galaxies with a spectroscopic redshift measurement and
a reliability flag $1 \leq flag \leq 9$, 915 are stars
and 113 are type I AGN, making this galaxy and AGN sample the largest
homogeneous sample with spectroscopic redshifts at this depth.
A redshift measurement has not been possible for
1315 objects, bringing the redshift measurement completeness of the full (Epoch 1 + Epoch 2)
VVDS-Deep sample to 89.5\%. 
The total area covered is 0.74 square degrees, the redshift range $0.0024 \leq z_{Deep} \leq 5$
for a mean redshift $\bar{z}=0.92$ (Le F\`evre et al., 2013), and the
volume sampled is $\sim0.4\times10^7$h$^{-3}$Mpc$^3$.

\subsection{VVDS Ultra-Deep}

\begin{figure}
   \centering
   \includegraphics[width=8cm]{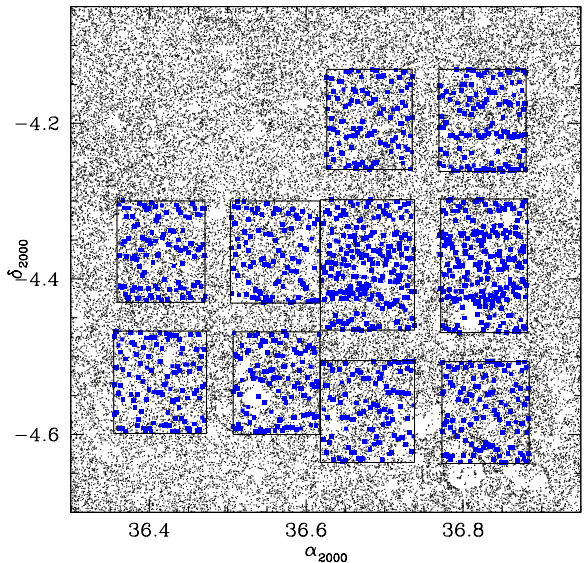}
      \caption{The distribution in ($\alpha, \delta$) of galaxies observed in the 
               Ultra-Deep $23 \leq i_{AB} \leq 24.75$ sample (heavy symbols) over the
               photometric parent sample (dots). 
              }
         \label{dist_udeep}
   \end{figure}

This latest component of the VVDS was aimed to produce an i-band
limited magnitude survey 2.25 magnitudes beyond the VVDS-Wide, 
and 0.75 magnitudes fainter than the VVDS-Deep, reaching $i_{AB}=24.75$. 
Very deep spectroscopic observations have been performed in service mode 
in the framework of ESO Large Program 177.A-0837, integrating 18h per target
in each of the LRBLUE and LRRED grisms for 3 VIMOS pointings. These cover a field of 512 arcmin$^2$ roughly centered
at $\alpha_{2000}=02h26m28.8s$ and $\delta_{2000}=-4^{\circ}23\arcmin06\arcsec$, 
included in the VVDS-02h Deep field  (0226-04) area, as shown in Figure \ref{dist_udeep} and listed in Table \ref{obs}. 

To reach the depth of $i_{AB}=24.75$ and ensure a high completeness in redshift
measurement, we have devised a strategy with
long integrations and an extended wavelength coverage using
VIMOS.
An essential component to keep 
a high completeness in measuring redshifts for $z>1$ objects
is to reduce redshift degeneracies by increasing the observed 
wavelength range and therefore maximizing the number of observed spectral features. 
We elected to combine VIMOS 
low-resolution blue grism observations over $3650 \leq \lambda \leq 6800$\AA
~and low-resolution red grism observations over $5500 \leq \lambda \leq 9350$\AA
~to cover a wavelength range 3650 to 9350\AA, the overlapping region from 5500 to 6800 \AA~
then getting a total exposure time of 36 hours. 

We have been designing slit masks from the extensive and 
deep multi-wavelength photometry from the CFHTLS, including three target samples:
{\it (i)} a randomly selected sample of galaxies with $23 \leq i_{AB} \leq 24.75$,
{\it (ii)} a sample of galaxies with $22.5 \leq i_{AB} \leq 24$ randomly
selected from galaxies with flags 0, 1 or 2, as measured
in the VVDS-Deep sample, and 
{\it (iii)} a sample of 'targets of opportunity' with
objects selected from GALEX Lyman-break (GLBGs) candidates at $z\sim1$ and 
extremely red objects (EROs) selected from their red colours and aimed at picking-up 
high redshift $z>1$ passively evolving red early-type galaxies. 
In the following we call these samples the 'Ultra-Deep', the 'Deep-re-observed', 
and the 'colour-selected' samples. The VMMPS mask-design software (Bottini
et al., 2005) was first used to force slits on the colour-selected sample (a few per VIMOS 
quadrant), then to force slits on the Deep-re-observed sample (about 20 per
quadrant), and finally to place slits randomly on the Ultra-Deep sample (about
80 per quadrant), for an average total number of about 450 slits observed in
one observation.

As described in Section \ref{zmeas}, the data have been processed using the VIPGI software.  
The redshift measurements
have been performed in several steps using the EZ engine.
The redshifts of the blue spectra and the red spectra have been
measured separately, each by 2 persons in the team, and two reconciled lists of 
redshifts derived from the blue and red observations have been
separately produced. The 1D blue and red digital spectrograms of each galaxy have
then been joined, 
and redshifts from these combined spectra have been
derived again separately by two persons, without knowledge of 
the redshifts derived from blue or red observations. The final redshifts were
then assigned by two persons jointly comparing the three
different redshift measurements (blue, red, and joined), and deciding on the 
associated reliability flags. We have added a new flag
'1.5' to indicate the spectroscopic redshifts with a low
reliability flag=1, but  which agree with the photometric redshifts 
to within $\delta z=0.05 \times (1+z)$ (see section \ref{pz}).

   \begin{table*}
      \caption[]{VVDS Ultra-Deep observations in the VVDS-02h field}
      \[
         \begin{array}{p{0.1\linewidth}lllcc}
           \hline \hline
            \noalign{\smallskip}
            Pointing      &  \alpha_{2000} & \delta_{2000} & $Grism$ & $Exposure $ & $Seeing $\\ 
                 &  &&  & $time (min)$ & \\ 
            \noalign{\smallskip}
            \hline
            \noalign{\smallskip}
            D02P001 &  02h27m00.72s & -04\deg16\arcmin45.3\arcsec & LRRED & 40\times27 & 1.2\arcsec \\ 
            D02P001 &  02h27m00.72s & -04\deg16\arcmin45.3\arcsec & LRBLUE & 40\times27 & 1.0\arcsec \\ 
            D02P002 &  02h27m01.30s & -04\deg29\arcmin12.0\arcsec & LRRED & 40\times27 & 1.2\arcsec  \\ 
            D02P002 &  02h27m01.30s & -04\deg29\arcmin12.0\arcsec & LRBLUE & 40\times27 & 1.1\arcsec \\ 
            D02P003 &  02h25m57.25s & -04\deg26\arcmin54.5\arcsec & LRRED & 40\times27 & 0.9\arcsec  \\ 
            D02P003 &  02h25m57.25s & -04\deg26\arcmin54.5\arcsec & LRBLUE & 40\times27 & 0.9\arcsec  \\ 
\noalign{\smallskip}
            \hline
         \end{array}
      \]
\label{obs}
   \end{table*}


\subsubsection{The Ultra-Deep $23 \leq i_{AB} \leq 24.75$ sample}

The main Ultra-Deep sample contains a total of 815 objects which have been  
observed: we have measured the redshifts for 721 galaxies, 3 type-I AGN, and 23 stars, 
673 of which have a flag larger or equal to 1.5,
representing $\sim83$\% of the sample. 
The total area covered is 0.14 square degrees, the redshift range is $0.05 \leq z_{Ultra-Deep} \leq 4.5$
for a mean redshift $\bar{z}=1.38$ (Le F\`evre et al., 2013), and the volume sampled is 
$0.5\times10^6$h$^{-3}$Mpc$^3$.



\subsubsection{The re-observed deep $17.5 \leq i_{AB} \leq 24$ sample}
\label{reobs}

In these deeper Ultra-Deep observations, we have re-observed  the galaxies in the original 
VVDS-Deep (Le F\`evre et al. 2005a) for which we failed
to get redshifts (flag 0), those with lower reliability (flag 1), 
and have re-observed a sample of galaxies with higher redshifts reliability (flag 2) 
to further assess their statistical robustness.
The main goal was to get secure spectroscopic 
redshifts for these objects, and hence obtain a statistical estimate of the true redshift
distribution of the galaxies in the VVDS-Deep which had lower redshift reliability flags. 
This sample has been randomly build  from the objects with flags
0, or with flags 1 and 2 and redshifts $z \geq 1.4$  in the Epoch-1 VVDS-Deep sample. 
This redshift $z=1.4$ marks the point when the [OII]3727\AA ~line starts to be difficult to
detect as the LRRED sensitivity and strong sky OH bands affect this line above $\sim9000$\AA, and 
at $z=1.5$ this line  is beyond our wavelength domain,
leaving only weak absorption features from the UV-rest spectrum in the observed domain.

A total of 241 objects have been successfully targeted. The deeper observations enable
to secure a large number of redshifts for these sources, 
with 153 objects with a very high reliability flag above 3, and 72 objects with 
a reliable flag from 1.5 to 2.5.
 
\begin{figure}
   \centering
   \includegraphics[width=8cm]{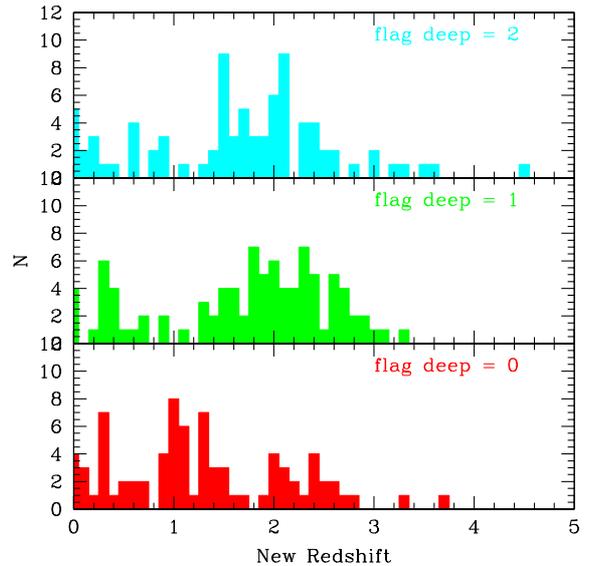}
      \caption{The redshift distribution of 
               objects with spectroscopic flags 0, 1 and 2 from the VVDS-Deep
               sample with $17.5 \leq i_{AB} \leq 24$, re-observed in the VVDS-UltraDeep survey.
              }
         \label{histz_re}
   \end{figure}

These new measurements enable to better understand the incompleteness
of the VVDS-Deep sample, as discussed in section  \ref{comp_deep}.
The redshift distribution of this sample, for each original VVDS-Deep flag, 
is presented in Figure \ref{histz_re}. The redshift distribution of 
VVDS-Deep galaxies with flag=0 shows that the redshift failures in VVDS-Deep
are coming from the full redshift range, although with a preference
for $1<z<2.5$ including the LRRED grism 'redshift desert' (see Section \ref{wave}).
The re-observed flag 1 and flag 2 are also preferentially in this 'redshift desert', 
as expected because the wavelength range of the LRRED grism 
makes it difficult to identify features in this difficult redshift range.  

Using this re-observed VVDS-Deep sample has enabled a statistical correction
of the full sample using the weighting scheme described in Section \ref{compl}.


%
%
%
%



\subsection{Serendipitous observations of Lyman Alpha Emitting galaxies}

As we were processing 2D spectra of the main targets of this program,
a number of objects with single emission lines have been identified
falling at random positions along the slits. This is not surprising given the depth 
of the survey, and, as the slits extend into blank sky areas, the high number of slits
implied a large serendipitous survey of a significant sky area.
For the VVDS-Deep observations of the 0226-04 field, the total sky area observed
through the  slits is 22 arcmin$^2$, while for the Ultra-Deep observations
in 3 masks the total sky area amounts to a 
total of 3.2 arcmin$^2$. The majority of these single emission
line objects have been identified as Lyman--$\alpha$ emitters
with redshifts $2 \leq z \leq 6.62$, as described in Cassata et al. (2011).

\subsection{Multi-wavelength  data in the VVDS surveys}
\label{ima}

The VVDS surveys have been conducted in fields with a large range of multi-wavelength
data, as summarized here.

The VVDS-Wide fields have been observed with the CFH12K camera at CFHT (Le F\`evre et al. 2004a),
reaching depth of $I_{AB}=24$ ($3\sigma$) (McCracken et al. 2003). 
All VVDS-Wide fields have BVRI photometry, except the 1400+05 which has BRI photometry.

The VVDS-02h 0226-04, including the Deep and Ultra-Deep surveys, 
has been the target of a number of multi-wavelength observations. 
Improving on the early CHF12K BVRI survey described above and U-band imaging (Radovich et al., 2004), 
the CFHT Legacy Survey (CFHTLS\footnote{See the data release and associated documentation at http://terapix.iap.fr/cplt/T0007/doc/T0007-doc.html}, Cuillandre et al. 2012) 
D1 field is including all of the VVDS-02h area observed in
the u'gri and z filters with the Megacam at CFHT, with seeing FWHM
from 0.75 to 0.99 and reaching 50\% completeness magnitude for point sources 
at depths of 26.96, 26.73, 26.34, 25.98, and 25.44 in these bands, respectively.
Following the initial survey of Iovino et al. (2005) and Temporin et al. (2008) in this field,
new near-infrared photometry has become available from the WIRDS survey (Bielby et al. 2012), 
reaching 50\% completeness for point-sources at $J_{AB}=24.90$, $H_{AB}=24.85$
and $Ks_{AB}=24.80$ with 0.80, 0.68, 0.73 arcsec seeing FWHM images respectively. 
Other multi-wavelength data are available in this field,
with GALEX (Arnouts et al. 2005), XMM (Pierre et al. 2004), Spitzer-SWIRE (Lonsdale et al. 2003), 
and VLA (Bondi et al. 2003).
This field has been the target of the Herschel HERMES survey (Oliver et al. 2012),
and matched to VVDS data (Lemaux et al. in preparation).

The VVDS-22h 2217+00 field has been receiving additional photometric observations
since the VVDS imaging completion. Near infrared imaging in J and K bands reaching 
$K=21$ (Vega) has been obtained by the UKIRT UKIDSS-DXS survey (Lawrence et al. 2007).
This field has been extensively imaged in u',g,r, i, and z bands 
by the CFHTLS over an extended area named CFHTLS-W4 covering a total of 
25 square degrees in a SE-NW oriented stripe.

All the multi-wavelength data has been cross-correlated with the VVDS spectroscopic 
sample and is made available on the VVDS database.

\section{Completeness and selection function}
\label{compl}

\subsection{Photometric redshifts of the full sample}
\label{pz}

\begin{figure}
   \centering
   \includegraphics[width=8cm]{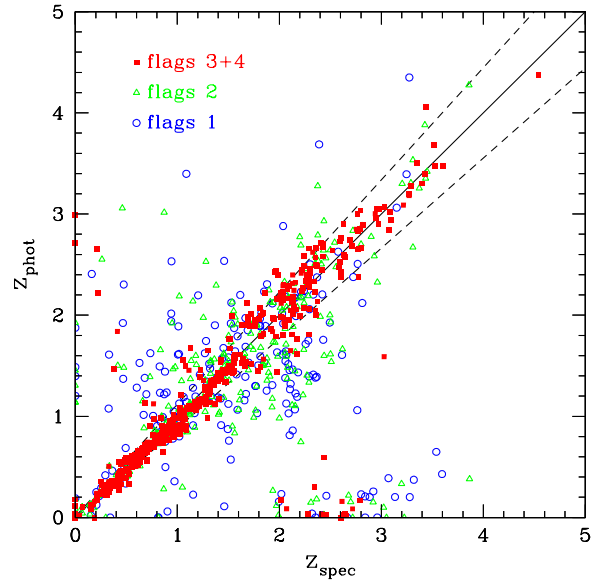}
      \caption{Comparison of photometric and spectroscopic redshifts,
              for different spectroscopic flags. The dashed lines represent agreement
              between the two within $|dz| \leq 0.05 \times (1+z)$. Filled (empty) symbols are for
	      galaxies with (without) JHK photometry, respectively.
              }
         \label{photoz}
   \end{figure}

To help understand the spectroscopic completeness, we are using
photometric redshifts computed following the method described in
\cite{ilbert06} and \cite{ilbert09}. 
We have used the CFHTLS data release v5.0, with u',g,r,i and z' 
photometry reaching 80\% magnitude completeness limits for point sources
of 26.4, 26.1, 25.6, 25.3, 25.0, respectively.
We have added near-IR photometry from the WIRDS survey in J H and Ks bands (Bielby et al., 2012). 
The photometric  redshifts have been trained on a spectroscopic sample including
the highly reliable (flags 3 and 4) spectroscopic redshifts 
from the VVDS-Deep (Le F\`evre et al. 2005a) 
and new spectroscopic redshifts with flags 3 and 4 from the Ultra-Deep
survey, following the method described in \cite{ilbert06}. The comparison of 
spectroscopic redshifts and photometric redshifts is shown
in Figure \ref{photoz}. 

We find that there is an excellent agreement between the spectroscopic
and photometric redshifts with $dz = 0.05 \times (1+z)$. The
rate of photometric redshifts catastrophic failures is below 3\% for all flags. 
There is a small number of catastrophic failures in the redshift range $2 \leq z \leq 3.5$,
including objects with very secure spectroscopic redshifts (flags 3 and 4),
this is due in part to objects for which the NIR photometry is lacking (e.g. on masked areas in the NIR
images).

\subsection{Sampling rates}
\label{selra}

The selection function of a spectroscopic redshift survey proceeds from
different steps in constructing the final sample, starting from
the photometric sample from which targets are selected.
To describe the completeness of the VVDS we make use of
the following definitions. $N_{phot}^{full}$ is the number of objects
in the full photometric catalogue limited by the magnitude range
of the survey: $17.5 \leq I_{AB} \leq 22.5$ for the Wide survey,
$17.5 \leq I_{AB} \leq 24$ for the Deep, and $23 \leq i_{AB}
\leq 24.75$ for the Ultra-Deep. 
$N_{phot}^{parent}$ is the number of objects in the
parent catalogue used for the spectroscopic target selection that
corresponds to the full photometric catalogue after removing the objects
already observed in this area.
$N_{target}$ is the number of targets selected for spectroscopic observations and
$N_{spec}$ is the number of objects for which one is able to measure a
spectroscopic redshift (using the flag system described in Section \ref{zmeas}). 

With this formalism, we can easily define the completeness of the VVDS 
spectroscopic samples as the combination of three different sampling rates,
as described below.

As the VVDS is purely magnitude selected, the first sampling to consider
is the fraction of objects targeted compared to the number of objects 
in the full photometric catalog, as a function of i-band magnitude. 
We call this the Target Sampling Rate (TSR) defined as the ratio
$N_{target}$/$N_{phot}^{parent}$.  This ratio does not depend on the magnitude
of the sources, as the VVDS targets
have been selected totally randomly within the parent photometric
catalogue.  The TSR varies with position on the sky, as described 
in Section \ref{masks}. The weight associated to the TSR is
$w_{TSR}=1/TSR$.

A second factor to take into account is the Spectroscopic Success Rate (SSR), the ratio $N_{spec}$/$N_{target}$.
It is a function of both the selection magnitude and redshift. To
determine the dependence of SSR on redshift, we use the spectroscopic
redshift for the targets that yield a  redshift, and the
photometric redshift for all the other targets.  
The weight associated to the SSR is $w_{SSR}=1/SSR$.

The third factor is the photometric sampling rate (PSR), computed as the ratio
$N_{phot}^{parent}$/$N_{phot}^{full}$. It applies only for the Ultra-Deep
survey as some of the brighter objects had already been observed on the Deep survey. The weight
associated to the PSR is $w_{PSR}=1/PSR$. For the Wide and Deep surveys $w_{PSR}=1$.



We expand below on the completeness computation for the VVDS-Deep sample,
as it can be further 
refined using the VVDS-Deep galaxies re-observed in the VVDS-UltraDeep. 
The TSR, defined as $N_{target}$/$N_{phot}^{full}$
($N_{phot}^{full}=N_{phot}^{parent}$ for VVDS-Deep),
is not strictly constant because the VVDS-Deep targeting strategy was
optimized to maximize the number of slits on the sky, slightly favouring the
selection of small objects, with a size significantly smaller than
the slit length. As a consequence, the TSR depends on the
x-radius of the objects, the projection of the
angular size of the objects on the slit (for further details we refer
the reader to Ilbert et al. 2005).
The SSR of the re-observed VVDS-Deep needs a more detailed treatment. 
It is defined as above ($N_{spec}$/$N_{target}$), but the redshift
distribution of objects with flag=1,2,9 is corrected 
using both photometric redshifts and the spectroscopic
redshifts of the objects that were re-observed with deeper integration
and larger wavelength coverage in the VVDS Ultra-Deep (see Sect. \ref{reobs}).  
We used the remeasured redshifts as follows. A sample of $\sim 80$ 
objects with flag=1 and $\sim 80$ with flag=2 in the original VVDS-Deep
were re-observed in the VVDS-UltraDeep, and 
furthermore these objects were selected
with spectroscopic redshift $z_{VVDS-Deep}\geq1.4$. We
computed the $n(z)$ distribution of these objects using the re-observed
redshift values, and we rescaled it to the total number of flag 1 and
2 objects (in the VVDS first+second Epoch data) with $z\geq 1.4$. We
did it separately for flag 1 and 2, and we call these distributions
$n_{1,\geq1.4}(z)$ and $n_{2,\geq1.4}(z)$. Then we used the
photometric redshifts (see Sect. \ref{pz}) to compute the $n(z)$ of
flag 1 and 2 objects with spectroscopic redshift $<1.4$, as these
are demonstrated to be secure to $i_{AB} \leq 24$ (e.g. Ilbert et al.,
2006). We call these
distributions $n_{1,<1.4}(z)$ and $n_{2,<1.4}(z)$. Summing the two new
$n(z)$ (for $z<1.4$ and $z \geq 1.4$) for each flag we obtain the total
redshift distributions for the two classes of objects, and we consider
them 100\% correct. We call them $n_{1,corr}(z)$ and
$n_{2,corr}(z)$. For each class of flag 1 and 2 we then obtain the
original $n_1(z)$ and $n_2(z)$, made using the original spectroscopic
redshifts, and the corrected ones $n_{1,corr}(z)$ and $n_{2,corr}(z)$, made using a combination of
remeasured spectroscopic redshifts and photometric redshifts.
We computed the corrected redshift distribution also for flag=9
objects ($n_{9,corr}(z)$). We do not have re-observed spectra for this
class of targets, so $n_{9,corr}(z)$ is simply their photometric
redshift distribution, as this is very robust in this redshift
range (see Section \ref{pz}, Ilbert et al. 2009, and the comparison to MASSIV redshifts in Section \ref{zmeas}). 
We call M$_{n(z),i}$ the ratio 
$n_i(z)$/$n_{i,corr}(z)$, where i=1,2,9
(according to the flag). We note that M$_{n(z)}$ is a modulation of
the original $n(z)$, so it does not change the total number of
objects. On the contrary, both the TSR and SSR  are needed
to take into account missed objects.
To derive the $N_{spec}$ and $N_{target}$ as a function of
redshift we weighted flag=1,2,9 objects by $w_{M,i}=1/M_{n(z),i}$
(where i=1,2,9) while we do not apply any weight to the counts of
flag=3 and 4 objects. To compute $N_{target}$ as a function of $z$ we
also need the $n(z)$ of flag=0 objects.  We use the $n(z)$ of the
re-observed flag=0 (see Sect. \ref{reobs}), normalized to the total
number of flag=0 in our sample. We finally computed the
$SSR=N_{spec}$/$N_{target}$ as a function of both magnitude and
redshift, using the remodulated $N_{spec}$ and $N_{target}$. 
It is important to note that, when using the $n(z)$ of photometric
redshifts, we corrected it for the failure rate in the determination
of photometric redshifts themselves. We computed the failure rate as
the ratio between the spectroscopic $n(z)$ of objects with flag=3 and
4, and their photometric $n(z)$ (Section \ref{pz}).

%

\subsection{Selection function for the Ultra-deep sample}

\begin{figure}
   \centering
   \includegraphics[width=8cm]{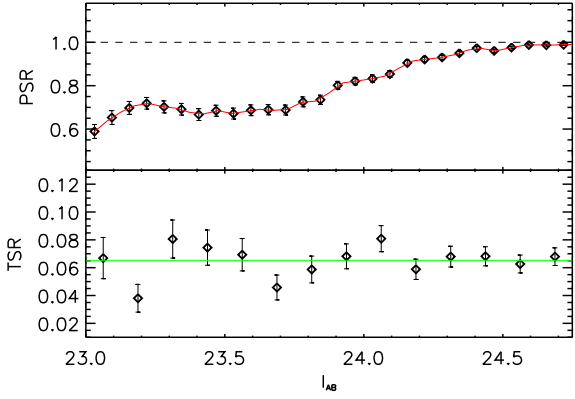}
      \caption{Photometric Sampling Rate (PSR, top) and Target Sampling Rate (TSR, bottom) vs. magnitude 
               for the redshift measurement 
               of the Ultra-Deep $23 \leq i_{AB} \leq 24.75$ sample.
              }
         \label{SRudeep}
   \end{figure}

\begin{figure}
   \centering
   \includegraphics[width=8cm]{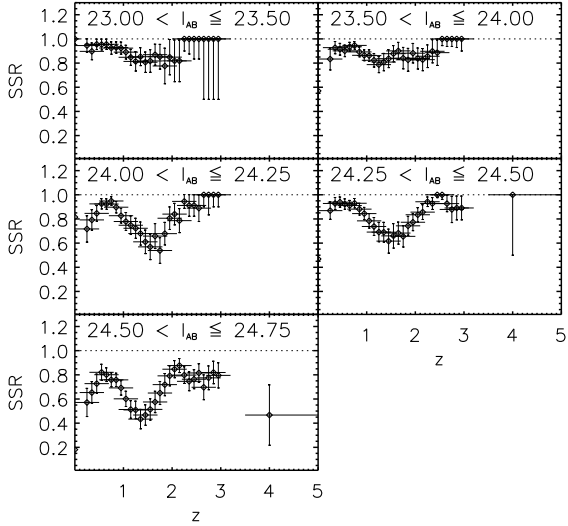}
      \caption{Spectroscopic Success Rate (SSR) for the redshift measurement of the Ultra-Deep $23 \leq i_{AB} \leq 24.75$
               sample, in bins of increasing magnitude, and as a function of redshift.
              }
         \label{SSRudeep}
   \end{figure}

The PSR, TSR and SSR for the Ultra-Deep sample are shown in Figures \ref{SRudeep}
and \ref{SSRudeep}. The PSR is rising with magnitude, starting at $PSR\sim0.7$ 
as the VVDS-Deep observations reduce the available number of 
targets, and reaching PSR=1 above $i_{AB}=24$, a magnitude range
where no galaxies had been observed in previous observing campaigns. 
The TSR is constant with magnitude, indicating that 6.5\% of objects with $23 \leq i_{AB} \leq 24.75$
have been observed in spectroscopy. 
The SSR is more complex: as magnitude increases the SSR gets globally lower 
and the success rate varies with redshift. At redshifts up to $z\sim1.2$ the SSR goes from 1 at the 
bright end to $\simeq0.8$ at the faint end, a possible result of the lack of emission features
in these objects. In the range $1.2<z<2$, the SSR is starting at
$SSR\sim0.9$ and is getting worse to $SSR\sim0.6$ at $z\sim1.5$ in the faintest magnitude bin. With
the large wavelength coverage observed, we can see that the 'redshift desert' is being 
crossed without much difficulty, but the range $1.5 \leq z \leq 2$ is still 
affected by some incompleteness. The redshift range above $z\sim2$ benefits from a higher 
SSR of $0.8<SSR<1$, the result
of the combined large wavelength coverage following the key spectral features 
at these redshifts; above $z=2.5$ the SSR gets down to $\sim0.8$ only in
the faintest magnitude bin. 

\begin{figure}
   \centering
   \includegraphics[width=8cm]{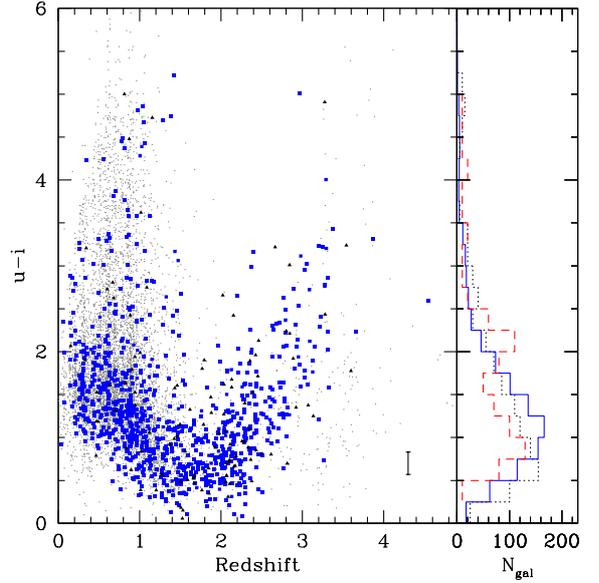}
      \caption{Observed $u-i$ colour distribution vs. redshift of the lowest reliability population (flags 1, black squares)
              compared to the distribution of galaxies with reliable redshifts (blue squares) 
              for the Ultra-Deep $23 \leq i_{AB} \leq 24.75$ sample. The distribution
              of colours vs. redshift in the Deep sample is shown in grey, and the typical $u-i$ error bar
              is shown on the lower right corner. The projected colour distribution
              is shown on the right panel as the blue histogram for reliable redshifts (2,3,4,9), the black dotted histogram
              for the lowest reliability population (flags 1) multiplied by 10 for comparison, and the red dashed histogram for the population
              of failed redshifts (flags 0) multiplied by 20 for comparison.
              }
         \label{fail_col}
   \end{figure}

Globally, the success rate in determining redshifts for this very faint sample is
quite high at 80\%, which ensures that no major population has escaped detection.
We show in Figure \ref{fail_col} the rest U-V colour distribution of the 
failed sample with flags 0 and 1: the colour distribution of the failed population
is not significantly different from the overall population.

\subsection{Selection function for the Deep sample}
\label{comp_deep}

\begin{figure}
   \centering
   \includegraphics[width=8cm]{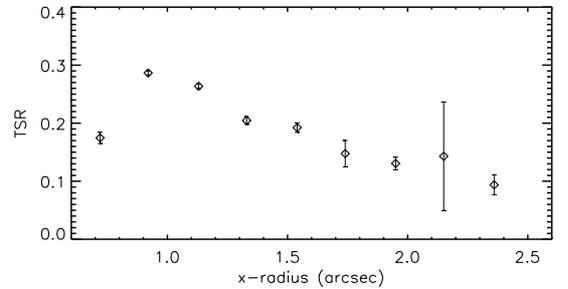}
      \caption{Target Sampling Rate (TSR) for the redshift measurement 
               of the Deep $17.5 \leq I_{AB} \leq 24$ sample, as a function
               of the projected size of objects along the slit.
              }
         \label{SRdeep}
   \end{figure}

\begin{figure}
   \centering
   \includegraphics[width=8cm]{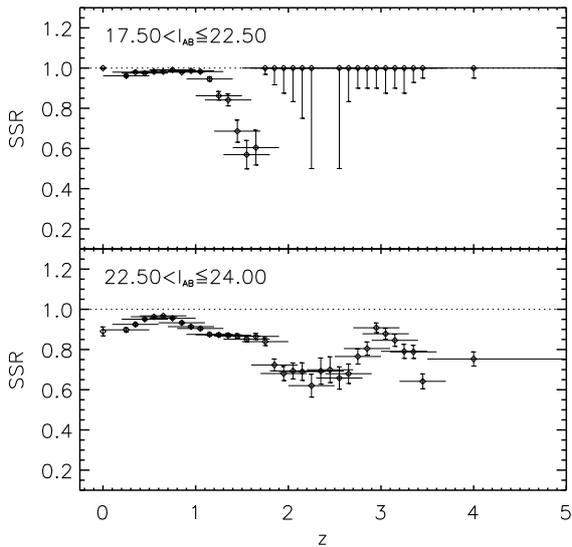}
      \caption{Spectroscopic Success Rate (SSR) for the redshift measurement of the Deep 
               $17.5 \leq I_{AB} \leq 24$ sample
              }
         \label{SSRdeep}
   \end{figure}

\begin{figure}
   \centering
   \includegraphics[width=8cm]{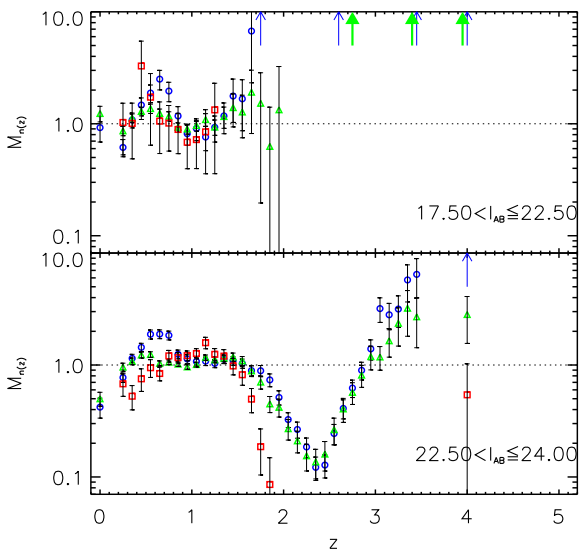}
	\caption{Modulation M$_{n(z)}$ of the redshift distribution 
                  of flag=1 (blue circles), flag=2 (green triangles)
                  and flag=9 (red squares) galaxies. See text for the
                  definition. The two panels represent two magnitude
                  ranges, as indicated in the labels. Arrows indicate
                  when the value of $M_{n(z)}$ is larger than 10: for
                  $z>1.7$ in $17.5\leq I_{AB} \leq 22.5$ and for
                  $z>3.5$ in $22.5\leq I_{AB} \leq 24$ for flag=1 (thin
                  blue arrows), and for $z>2.7$ in $17.5\leq I_{AB}
                  \leq 22.5$ for flag=2 galaxies (thick green
                  arrows). We do not have flag=9 galaxies for $z>1.5$
                  in $17.5\leq I_{AB} \leq 22.5$ and for $2<z<3.5$ in
                  $22.5\leq I_{AB} \leq 24.$.
              }
         \label{wdeep}
   \end{figure}

The TSR and SSR for the Deep sample are shown in Figures \ref{SRdeep}
and \ref{SSRdeep}. The PSR is irrelevant (PSR=1) for this sample. The TSR is 
ranging from 0.13 to 0.29 with a mean of 0.261, reflecting the areas where we observed
1, 2 or 4 times with VIMOS, and is varying with the radius of the object
along the slit as shown in Figure \ref{SRdeep}. The SSR is varying as a function
of magnitude and redshift (Figure \ref{SSRdeep}). For the faintest magnitudes
it is ranging from 0.92 at $z\sim1$ to 0.7 at $z\sim2.3$. In addition, we have 
corrected for the variation of the flag reliability with redshift, computing the 
weight $w_{129}$
as the ratio of the galaxies with the lower reliability flags 1, 2 and 9 over the
number of galaxies with photometric redshifts and re-observed spectroscopic redshifts in the same redshift bin,
as shown in figure \ref{wdeep}. While this ratio is constant at 1 up to $z\sim1.6$,
independently of magnitude, it is significantly lower than one in the range
$1.6 < z < 2.7$ and becomes higher than one for $z > 2.7$, correcting for the
fact that spectroscopic redshifts with low reliability flags have been assigned preferentially
at $z>2.7$ than in the redshift desert $1.6<z<2.7$. 


\subsection{Effect of wavelength domain on redshift completeness}
\label{wave}

\begin{figure}
   \centering
   \includegraphics[width=8cm]{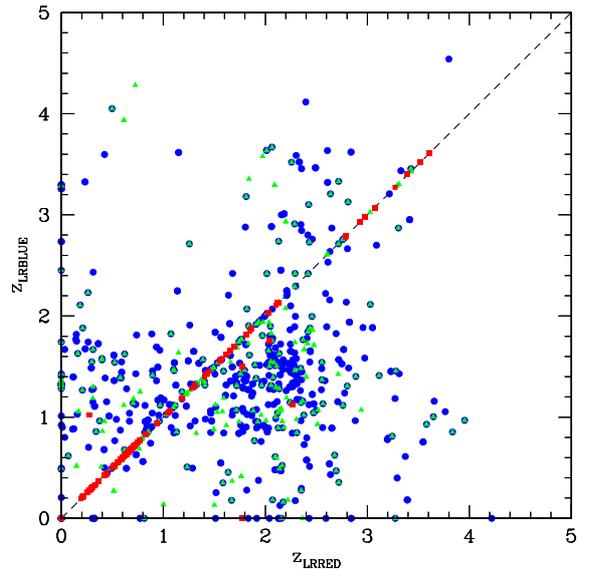}
      \caption{Comparison of redshifts measured from the LRBLUE
               3650--6800\AA ~VIMOS grism with redshifts obtained using the
               LRRED 5500--9350\AA ~grism (red squares: flags 3, 4 and 9; green triangles: flag 2;
               blue circles: flags 1)
              }
         \label{lrblue_lrred}
   \end{figure}

\begin{figure}
   \centering
   \includegraphics[width=8cm]{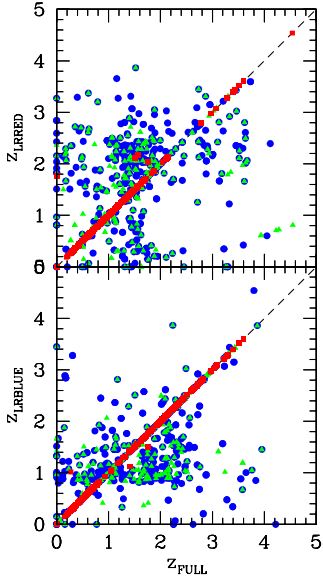}
      \caption{Comparison of redshifts measured from the LRBLUE
               3650--6800\AA ~VIMOS grism (bottom) or from the 
               LRRED 5500--9350\AA ~grism (top) with the best redshifts 
               using the full LRBLUE+LRRED 3650-9350\AA ~wavelength
               domain (red squares: flags 3, 4 and 9; green triangles: flag 2;
               blue circles: flags 1). 
              }
         \label{lrb_lrr_final}
   \end{figure}

\begin{figure}
   \centering
   \includegraphics[width=8cm]{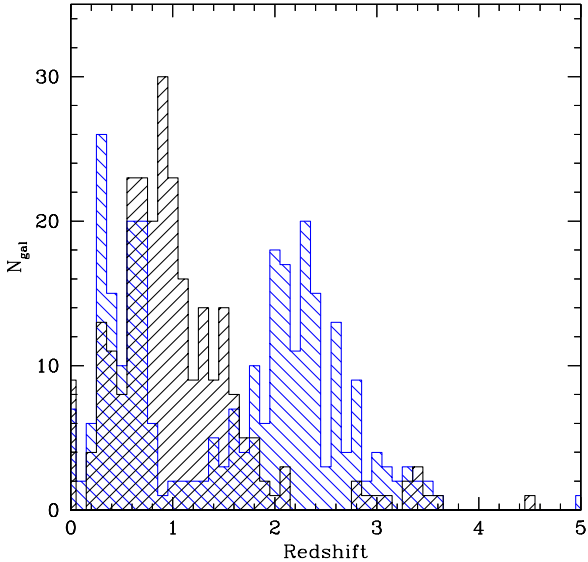}
      \caption{Redshift distribution of galaxies with the most reliable redshifts (flags 3 and 4),   
               from the LRBLUE setup (hatched blue histogram), and from the LRRED setup (hatched black histogram). 
              }
         \label{zblue_zred}
   \end{figure}

In the complex spectroscopic redshift measurement process, the observed wavelength
domain plays a critical role. In order to evaluate the impact of the 
wavelength coverage on redshift measurements we compare in
Figure \ref{lrblue_lrred} the redshifts 
obtained using either the blue (LRBLUE, 3650--6800\AA) or red (LRRED, 5500--9350\AA) 
VIMOS setups, and in Figure \ref{lrb_lrr_final}
we show the comparison between redshifts derived from the VIMOS blue or red setups and the 
final galaxy redshifts obtained when using the full wavelength domain 3650--9350\AA.
It is clearly seen that with the red setup alone, there is a trend for galaxies with low reliability
flags with $z>1.5$ to have their redshifts overestimated. This is directly related 
to [OII]3727 leaving the observed wavelength domain, leaving only weak features
as redshift increases until the Ly$\alpha$ line would enter this wavelength domain at
$z>3.5$. On the other hand, using the blue setup alone, galaxies with low reliability flags
and redshifts in the range $1.5 \leq z \leq 2.5$, have underestimated redshifts.
We can also see very well that combining the blue and red wavelength observations
to expand the wavelength coverage is needed to
cross the 'red redshift desert' at $2.2 \leq z \leq 2.8$ produced by the
red wavelength coverage missing out on the 1215--1900\AA ~rest frame
domain rich in spectral features (Ly$\alpha$, CIV-1549, etc.), 
and the 'blue redshift desert' at $0.8 \leq z \leq 1.5$
produced by the blue wavelength coverage corresponding to the 
absence of the 3727--4000\AA ~domain (with [OII]3727, CaH+K, 4000\AA ~break, etc.). 
This is best seen in comparing the redshift distributions of the
most reliable redshift measurements (flags 3 and 4) for each setup, as
shown in Figure \ref{zblue_zred}, where the complementarity of these
two wavelength domains is fully evident.

This experience of performing the observations of the same 
galaxies with a blue and a red setup with the
VIMOS spectrograph and independently measuring redshifts with
blue, red or a full wavelength coverage 3650--9350\AA
~therefore clearly demonstrates the benefit of using a large 
range covering most of the visible domain.
It is clear from our analysis that surveys 
using only a partial wavelength coverage in the 0.35--1 micron domain, like the DEEP2 in $6500-9100$\AA, the
VVDS-Deep and VVDS-Wide using $5500-9300$\AA ~(this paper) or the zCOSMOS-faint using $3600-6800$\AA ~(Lilly et al. 2007),
might be subject to  observational biases that must be carefully evaluated  
in terms of their spectroscopic success rate varying
with magnitude but also with redshift, as we have discussed in previous sections for the VVDS surveys.

\subsection{Spectroscopic and photometric masks}
\label{masks}

The selection of targets in the ($\alpha, \delta$) plane proceeds from the
combination of the geometric constraints from the photometric catalogs
and the placement of slits in the slit mask-making process. The knowledge of
the spatial selection of targets is an important part of the selection
function, e.g. for clustering analysis (de la Torre et al. 2011)
or group finding (Cucciati et al. 2010).

The photometric catalogs are carefully screened to identify regions where the photometry
is potentially affected, e.g. by bright stars and their halos, satellite 
trails, detector defects, and this information is stored in photometric region files.
The spectroscopic slit-masks design leads to geometric constraints (see Le F\`evre et al. 2005a)
which, depending on the number of observations at a given sky location, will create 
a TSR varying with ($\alpha, \delta$). This is stored in spectroscopic region
files indicating the $TSR(\alpha, \delta$).

The photometric and spectroscopic region files are made available as part of the
final VVDS release.

\subsection{Correcting for the selection function}

The VVDS sample can be corrected for the selection function described in previous
sections, counting each galaxy with the following weight $W_{gal,i}$:

$W_{gal,i}=  1 / _{PSR} \times 1 / _{TSR} \times 1 / _{SSR} \times 1 / w_{129}$

with $PSR$, $TSR$, $SSR$, and $w_{129}$ as described in Section \ref{selra}.
This provides corrected counts, and therefore forms the baseline for volume--complete
analysis in the VVDS. The luminosity function and star-formation rate evolution
presented in Cucciati et al. (2012), has used the latest VVDS weights as presented here.
The redshift distribution of magnitude limited samples at various depth, 
corrected from selection effects, is presented and discussed in Le F\`evre et al.  (2013).


\section{Properties of galaxies in the VVDS}
\label{sample}

\subsection{The final VVDS sample}

The observations presented in this paper complete the VIMOS VLT Deep Survey.
In all, the redshifts of a purely I-band magnitude limited sample 
of 35\,016 extragalactic sources have been obtained, including the redshifts of 34\,594 galaxies and 422 type I AGNs, 
as described in Table \ref{vvds}. 
The number of galaxies in several redshift ranges are provided in Table \ref{znum}. We note
the large redshift coverage of the VVDS, with redshifts of galaxies ranging from $z=0.0024$ to $z=6.62$,
and type-I AGN from $z=0.0224$ to $z=5.0163$. In particular,
the VVDS has assembled an unprecedented number of 933 galaxies with spectroscopic redshifts through the 'redshift desert' $1.5 < z < 2.5$.

In addition to the extragalactic population, a total of 12\,430 
galactic stars have an observed spectrum. This results from the deliberate absence of photometric star-galaxy separation
ahead of the spectroscopic observations to avoid removing compact extragalactic objects.
While galactic stars represent a high 34\% fraction of the wide survey, they represent only
8.2\% and 3.2\% of deep and Ultra-Deep samples, respectively, a modest cost to pay to be able
to retain all AGNs and compact galaxies in our sample.

\subsection{NIR--selected samples with spectroscopic measurements 
complete to $J_{AB}=23$, $H_{AB}=22.5$ and $Ks_{AB}=22$}
\label{irselect}

\begin{figure*}
   \centering
   \includegraphics[width=10cm]{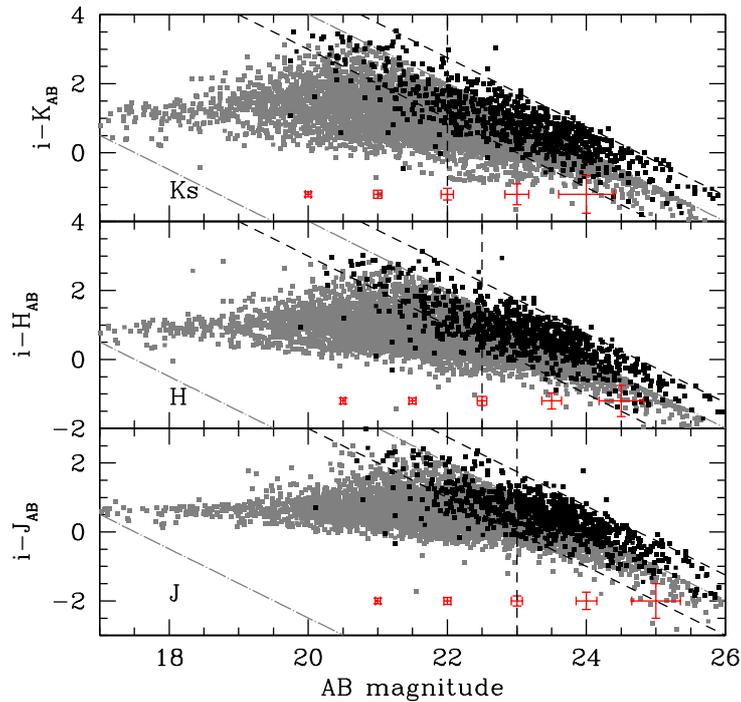}
       \caption{NIR colour magnitude diagram for J, H and Ks selected samples obtained from 
               the Ultra-Deep sample (black points): 
               The two oblique dashed lines correspond to the I-band limiting magnitudes
               of the survey $i_{AB}=23$ on the bright end, $i_{AB}=24.75$
               on the faint end. The VVDS-Deep colour-magnitude distribution is shown as the grey 
               points with the $i_{AB}=17.5$ and $i_{AB}=24$ limits indicated as long-dashed lines.
               We are able to extract samples nearly complete
               in spectroscopic redshift measurements, selected down to 
               $J_{AB}=23$, $H_{AB}=22.5$, and $Ks_{AB}=22$ (vertical dashed lines).
               The faint i-band magnitude cut imposes
               a loss of red objects for fainter magnitudes but still 80\%
               complete down to  $J_{AB}=24$, $H_{AB}=23.5$, and $Ks_{AB}=23$.  
               Typical error bars are shown at the bottom of each panel.
              }
         \label{ik_k}
   \end{figure*}

With the depth of the Ultra-deep sample to $i_{AB}=24.75$ and the
distribution of $I-J$, $I-H$ and $I-K$ colours, we are able to build
samples nearly complete  in spectroscopic redshift measurements
with 5\,846, 5\,207, and 4\,690 galaxies with $J_{AB} \leq 23$, 
$H_{AB} \leq 22.5$, and $Ks_{AB} \leq 22$, respectively, as shown in Figure \ref{ik_k}.
At these limits only a few percent of the reddest galaxies (e.g. with $i-Ks_{AB}>3$)
would escape detection. 
Fainter than this limit, the completeness is still 80\%
at $Ks_{AB}=23$ loosing only the redder objects with $i-Ks_{AB} \geq 2$,
and extends down to $Ks_{AB}=24$ for star forming objects
with $i-Ks_{AB} \leq 1$.



\subsection{Magnitude-redshift distributions}

\begin{figure*}
   \centering
   \includegraphics[width=8cm]{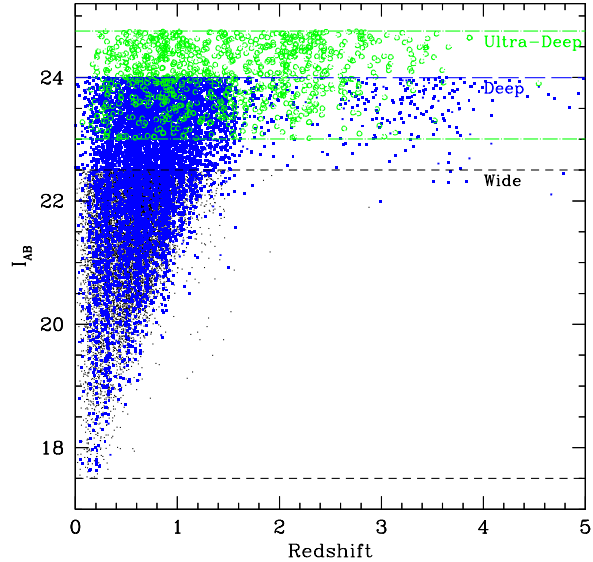}
      \caption{Distribution of apparent $I_{AB}$ apparent magnitudes for the 
               the VVDS-Wide (dots), VVDS-Deep (squares) and Ultra-Deep (open circles) samples.
              }
         \label{magi_z}
   \end{figure*}

The apparent $I_{AB}$ magnitude as a function of redshift for the 
VVDS Wide, Deep and Ultra-Deep samples shows the imposed magnitude
limits as presented in Figure \ref{magi_z}. The complementarity of
the three samples is evident, with the bright part of the luminosity
function best sampled in the Wide and Deep samples, the deep and
Ultra-Deep samples sampling the faint population, and the Ultra-Deep sample being mostly
unaffected by any instrument-imposed redshift desert.

\begin{figure*}
   \centering
   \includegraphics[width=10cm]{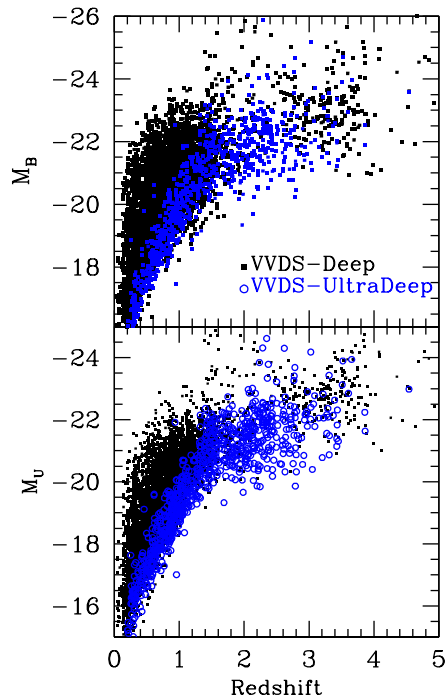}
      \caption{Distribution of B-band (top) and u-band (bottom)  absolute magnitudes for 
               the VVDS-Deep (dots) and Ultra-Deep (open circles) samples.
              }
         \label{abs_u}
   \end{figure*}

The distribution of absolute rest-frame u-band and B-band magnitudes
in the Deep and Ultra-Deep samples are presented in Figure \ref{abs_u}.
At redshift $z\sim1$ the VVDS surveys span a luminosity range
from $M_B\sim-18$ to $M_B\sim-23$, corresponding to 0.07 up to more than $7 \times$ 
the characteristic luminosity $M^*$ (Ilbert et al. 2005, Cucciati et al. 2012).

\subsection{Colour - magnitude evolution}

\begin{figure*}
   \centering
   \includegraphics[width=13cm]{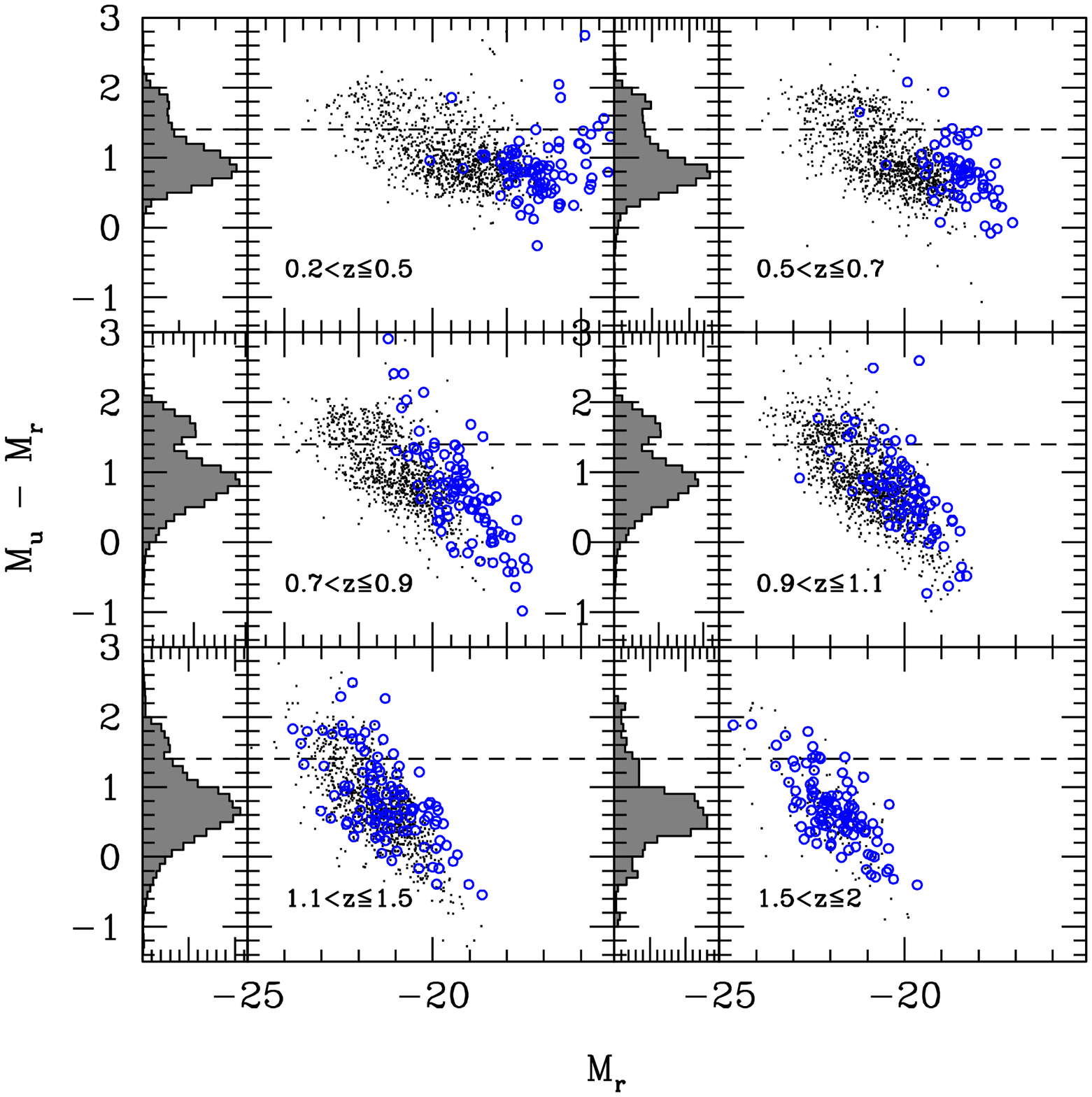}
       \caption{$M_u - M_r$ vs. $M_r$ colour-magnitude diagram for the VVDS-Deep (dots) and VVDS-UltraDeep (circles). A  
                colour bimodality is observed up to $1<z<1.5$.
              }
         \label{cmr}
   \end{figure*}

The $M_u-M_r$ vs. $M_r$ colour magnitude diagrams for the VVDS-Deep and Ultra-Deep
samples are presented as a function of redshift in Figure \ref{cmr}. 
As already noted from our earlier sample by Franzetti et al. (2007), the VVDS
data show a clear bimodality in colour, already present from redshifts
$1.5 < z < 2$, with a second peak in $M_u-M_r$ starting to be prominent at $1 < z \leq 1.5$ and below.
The VVDS traces up to $z\sim2$ the 'red sequence' originally identified in Bell et al. (2004)
up to $z\sim1$. 

\subsection{Average spectral properties}

\begin{figure*}
   \centering
   \includegraphics[width=18cm]{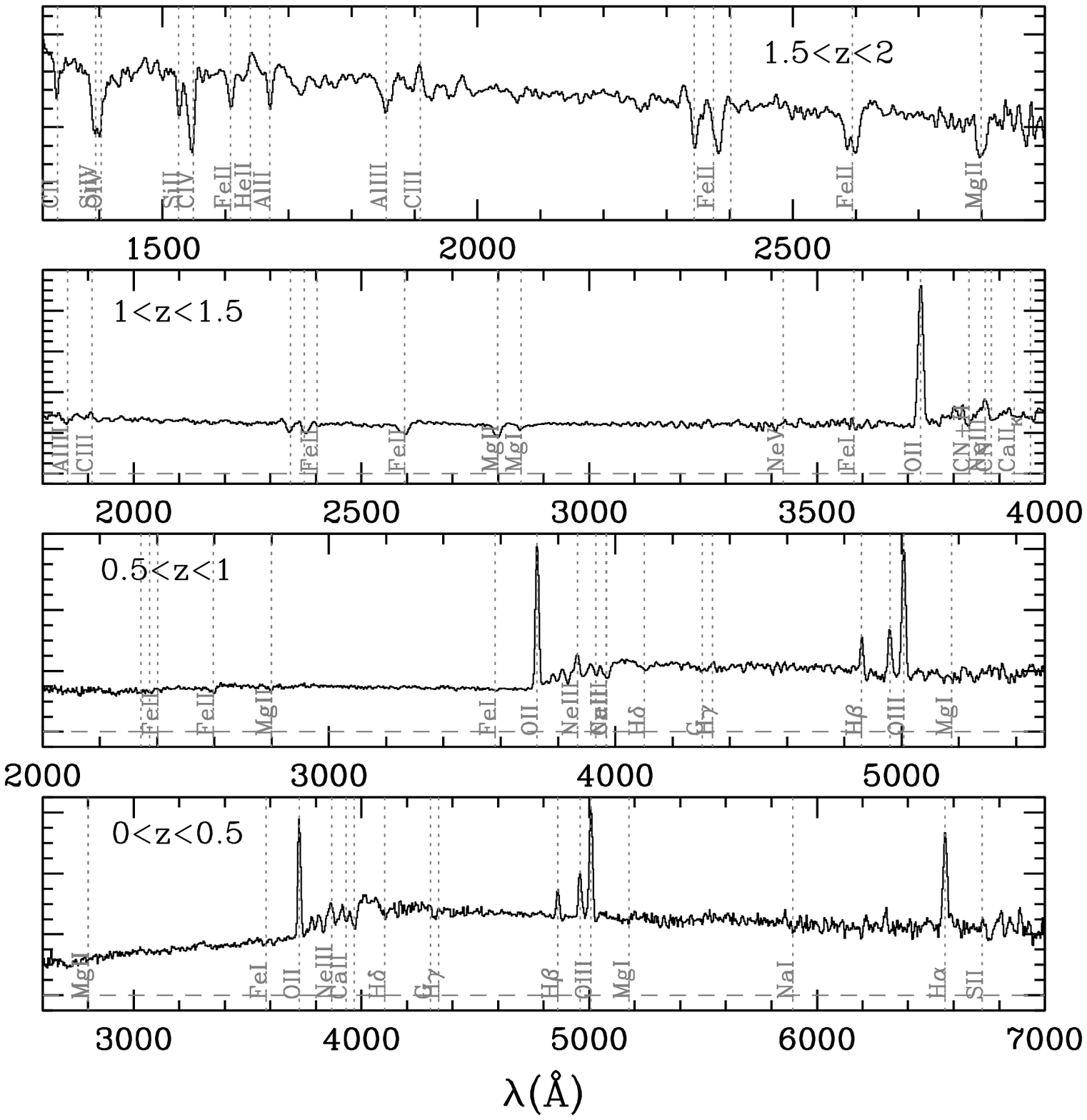}
       \caption{Average rest-frame spectra ($F_{\lambda}$) of all galaxies with flags 3 and 4 in the VVDS per redshift bin up to $z=2$.
              }
         \label{avg_spec1}
   \end{figure*}

\begin{figure*}
   \centering
   \includegraphics[width=18cm]{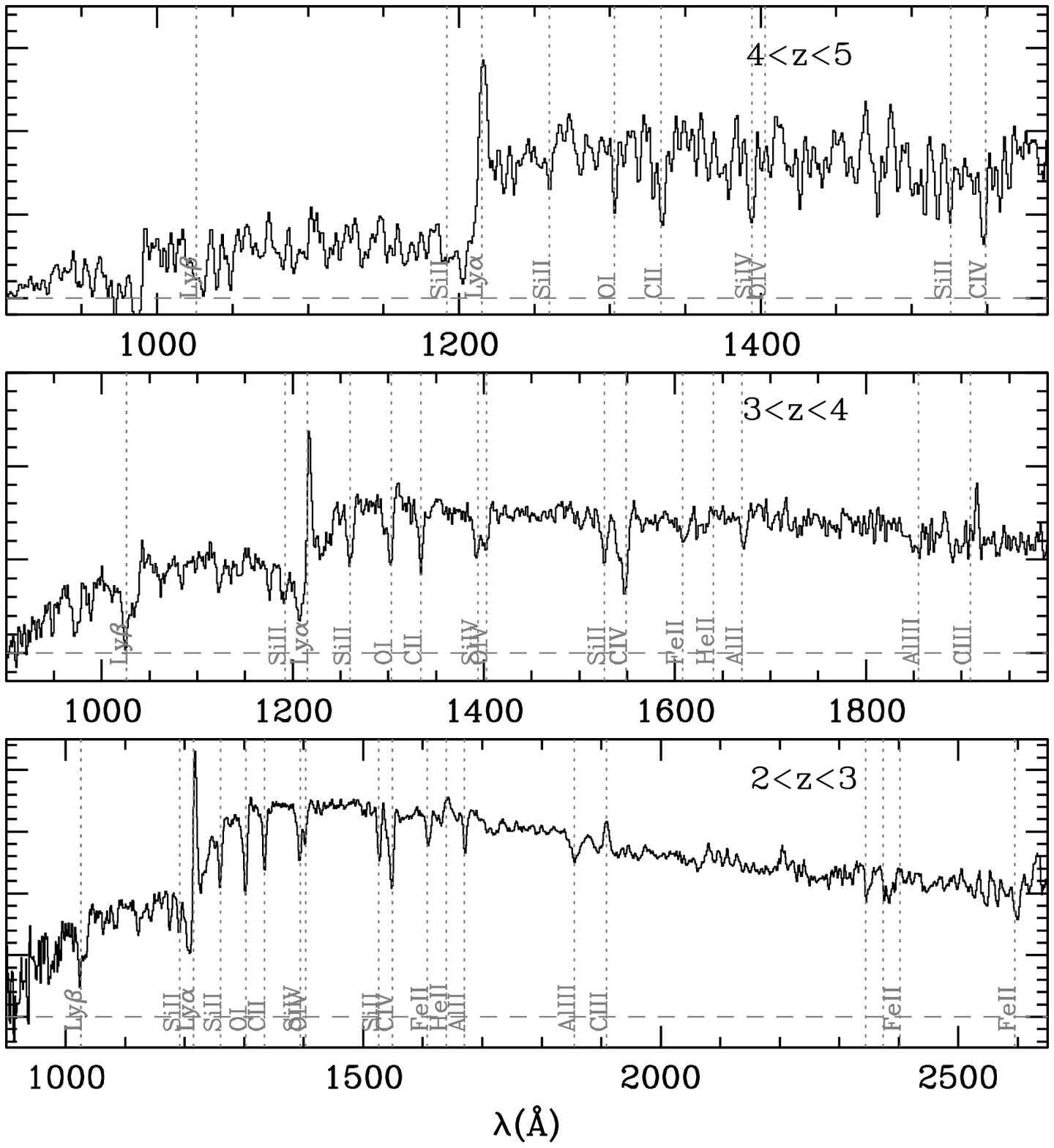}
       \caption{Average rest-frame spectra ($F_{\lambda}$) of all galaxies with flags 2, 3 and 4 in the VVDS per redshift bin from $z=2$ to $z=5$.
              }
         \label{avg_spec2}
   \end{figure*}

The average spectra of galaxies in the VVDS are presented in Figures \ref{avg_spec1}
and \ref{avg_spec2}. We have used the \texttt{odcombine} task in IRAF, 
averaging spectra after moderate sigma clipping and scaling to the same median continuum value,
and applying equal weight to all spectra. 

These spectra average over the range of spectral types, which cover from early-type 
to star-forming galaxies.  

\section{Comparison with other spectroscopic surveys at high redshifts}
\label{comparison}


Comparing different spectroscopic redshift surveys is not as straightforward as it may seem, 
as the parameter space defining them is quite large, and the science goals
may significantly differ.
Different flavours of multi-object spectrographs (MOS) enable different types of surveys, covering different
parts of the observing and science parameter space, hence special care
must be taken when comparing surveys. 
The most important survey output parameters are the number of galaxies observed, the limiting
magnitude, the sky area or volume covered and the redshift range surveyed. 
These are directly related to the technical performances of each MOS, 
including the instrument field of view, the wavelength coverage, the 
throughput, the number of simultaneous objects that can be observed (multiplex)
and the spectral resolution. In addition, the sample selection function is obviously 
of fundamental importance. If the goals of the survey
are more oriented towards galaxy evolution, then a proper sampling of the luminosity
and/or mass functions is of primary importance, while for surveys more oriented towards
large scale structures and cosmological parameters a fair sampling of all scales in large volumes is a prime concern.
In this context, attempts to rationalize a survey
performance using information theory, counting the number of bits of information, 
are largely illusory and it is much more useful to seek complementarity between surveys 
in a multi--dimensional performance space. 

\begin{figure*}
   \centering
   \includegraphics[width=20cm]{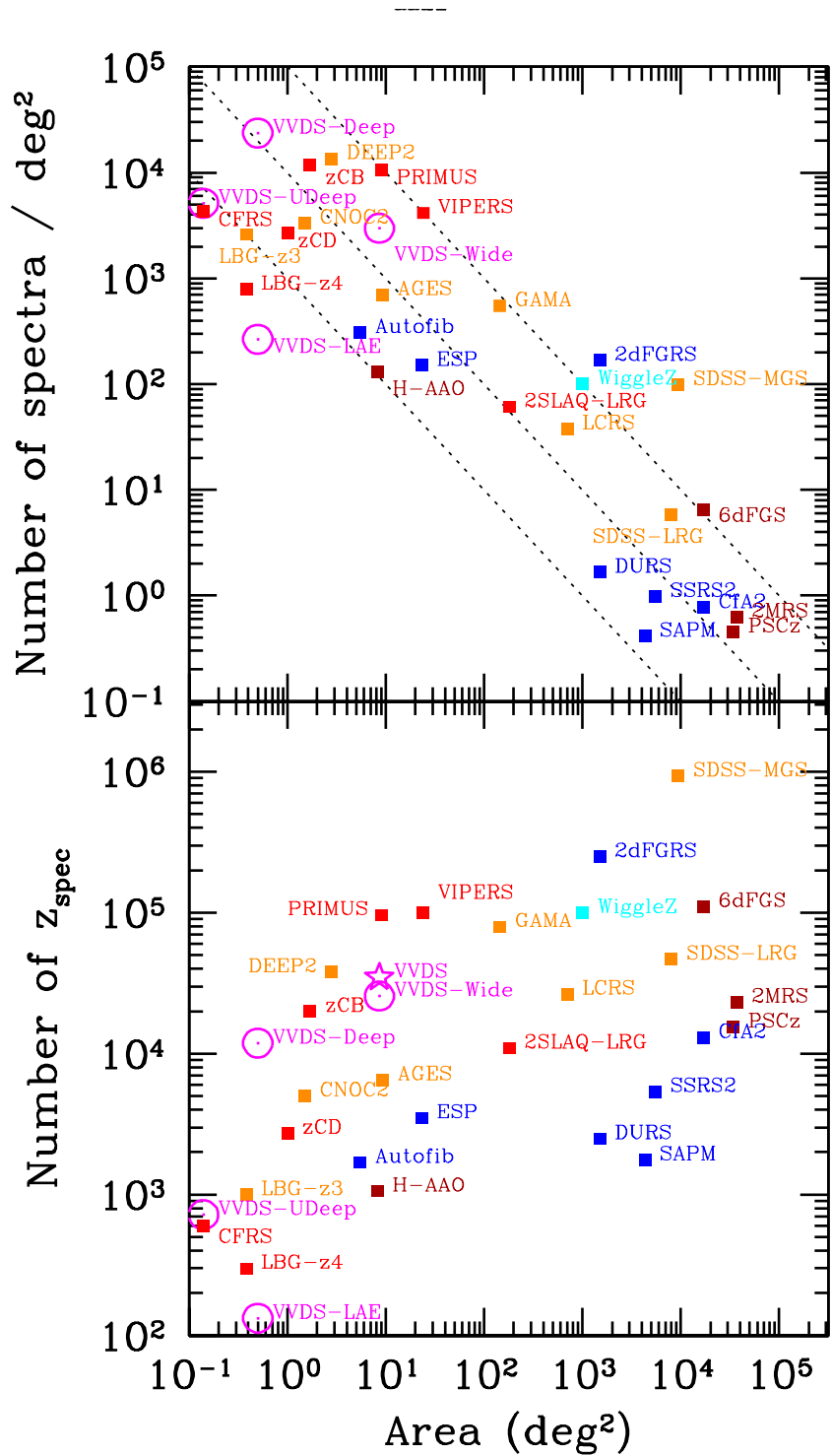}
      \caption{Comparison of the density of spectra (top) and the number of
               measured spectroscopic redshifts (bottom) versus the observed area, 
               for the VVDS (magenta) and different spectroscopic redshift surveys. The 
               combined VVDS surveys (Wide, deep and Ultra-Deep) are represented
               by the star symbol (bottom panel). 
               The colour code indicates the wavelength of the sample selection: UV (cyan), 
               B-band (blue), r-band (orange), 
               i-band (red, magenta), IR (brown). zCOSMOS-Bright and zCOSMOS-Deep are
               labelled 'zCB' and 'zCD', respectively. Dashed lines on the top panel represent
               samples with 1\,000, 10\,000 and 100\,000 redshifts (from bottom to top).
              }
         \label{dens_area}
   \end{figure*}

Following Baldry et al. (2010), we have compiled a list of spectroscopic surveys, as listed
in Table \ref{comp_surveys}, updated with new and on-going surveys, limiting to
spectroscopic samples larger than 100 galaxies.
In describing the GAMA survey, Baldry et al. (2010) have compared their survey to other
surveys in a plane relating the density of spectra to the area covered by each survey.
We present an updated version of this comparison in Figure \ref{dens_area}
(top panel), with the VVDS surveys as summarized in this paper. 
In this plane, the surveys distribution is somewhat bimodal, with larger area surveys with a lower
density of spectra on the bottom-right of the plot, and smaller area surveys with a
higher density of spectra in the upper-left. All high redshift survey (say beyond $z\sim0.5$)
are of this later category, with the VVDS-Deep presenting the highest density
of spectra, followed by the DEEP2 survey.  
The VVDS-Wide covers an area larger than zCOSMOS-Bright (zCB on the plot) or DEEP2,
at a lower spectra density. 
We show in Figure \ref{dens_area} (bottom panel) 
the number of spectroscopic redshifts obtained as a function
of area for these different surveys. The total number of spectra in the 
combined VVDS surveys is comparable to the DEEP2 survey, but lower than
the number of spectra in the PRIMUS (Coil et al., 2011) and VIPERS surveys limited to $z\sim1$.

\begin{figure*}
   \centering
   \includegraphics[width=20cm]{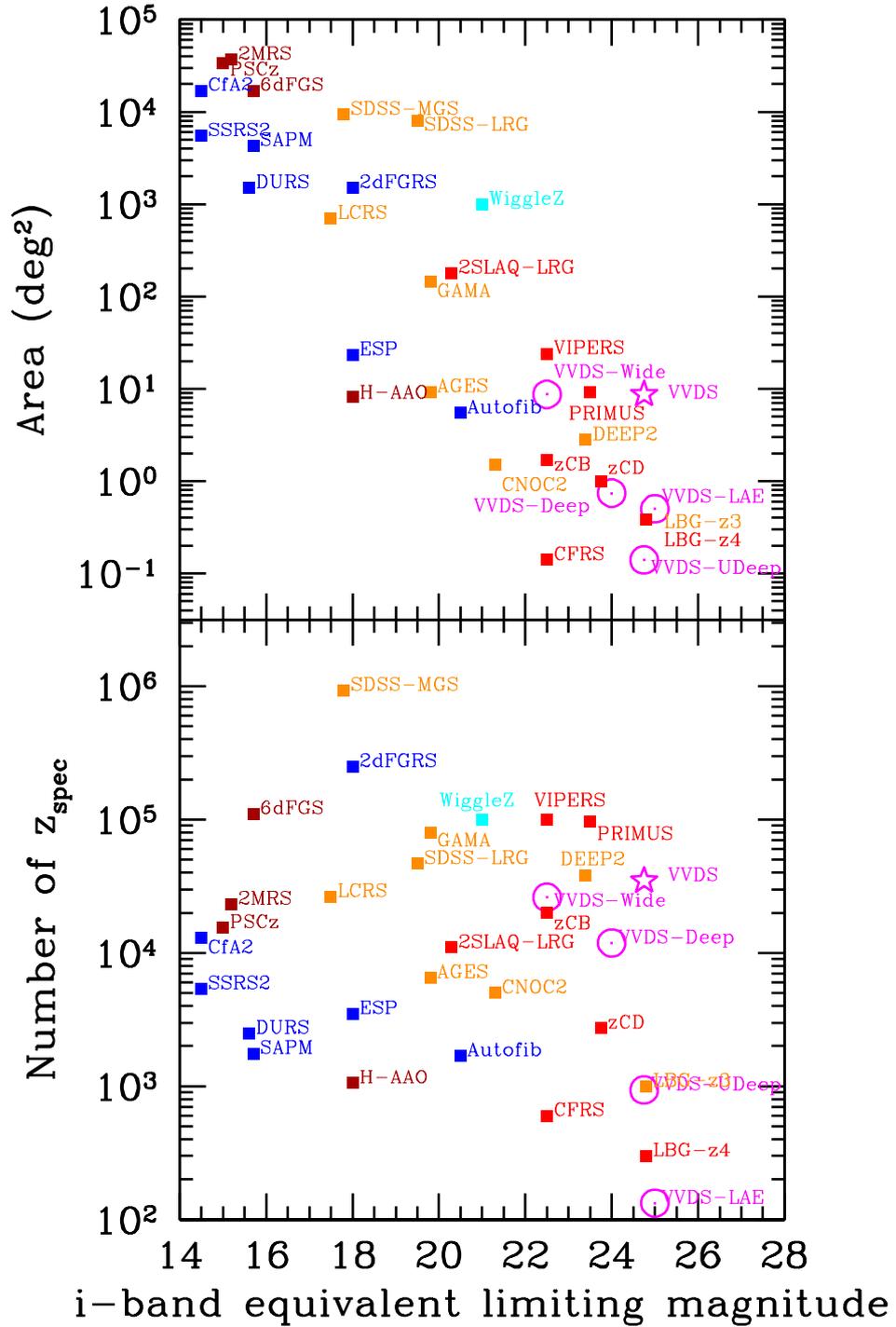}
      \caption{Comparison of the covered area (top)  and the number of measured
               spectroscopic redshifts (bottom) versus depth expressed as the equivalent i--band limiting magnitude, 
               between the VVDS (magenta) and different spectroscopic redshift surveys. The i--band limiting 
               magnitude of each survey has been estimated using the survey band and limiting magnitude
               in that band, and assuming a flat spectrum. The same colour code as in
               Figure \ref{dens_area} has been used.
              }
         \label{nspec_area_mag}
   \end{figure*}

Absent from the two panels in Figure \ref{dens_area} is the 
key information on the depth and the redshift coverage of each survey. 
For instance in these plots the VVDS-UltraDeep looks similar to
the CFRS as it covers about the same area with the same number density
and the same number of spectra, 
but while the CFRS extends to $z\sim1.2$ with $i_{AB} \leq 22.5$, 
the VVDS-UltraDeep extends to $z\sim4$ with $i_{AB} \leq 24.75$.
We therefore compare the number of spectra and covered area versus limiting magnitude 
in Figure \ref{nspec_area_mag}. 
The area (volume) covered by redshift surveys, as well as the number of spectra, 
are, as expected, getting smaller with increasing limiting magnitude or redshifts. 
The VVDS-Deep, VVDS-UltraDeep
and VVDS-LAE contribute some of the deepest spectroscopic redshift surveys to date.

\begin{figure*}
   \centering
   \includegraphics[width=20cm]{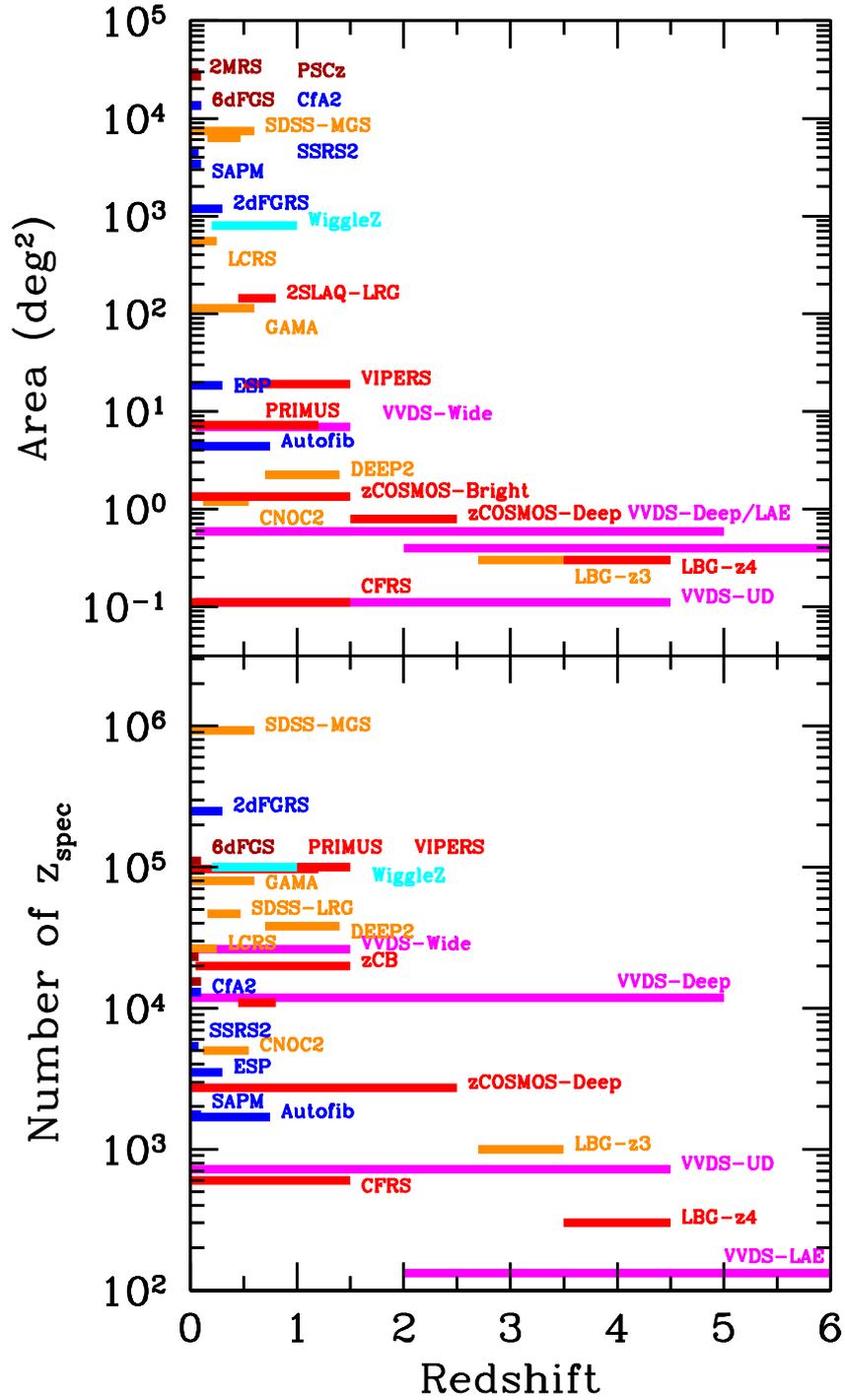}
      \caption{Comparison of the covered area (top)  and the number of measured
               spectroscopic redshifts (bottom) versus redshift, 
               between the VVDS (magenta) and different spectroscopic redshift surveys. The same colour code as in
               Figure \ref{dens_area} has been used
              }
         \label{nspec_area_z}
   \end{figure*}

To give further indications on the useful range
of these surveys, we compare in Figure \ref{nspec_area_z}
the number of redshifts obtained by the VVDS to other surveys, as well as
the area covered, as a function of redshift.  The uniqueness of VVDS is evident
as the VVDS surveys covers the largest redshift range of all surveys, and 
reach among the highest redshifts. 

Another interesting comparison between surveys is the total time necessary
to complete a survey. The observing efficiency of a survey
is characterized by the number of (clear) observing nights (hours) necessary to assemble the redshifts.
The VVDS-Wide, Deep, and Ultra-Deep surveys took 120h, 160h, and 150h on VIMOS at the VLT, respectively,
including all overheads.
By comparison at $z\sim1$, the CFRS took the equivalent of 19 clear nights ($\sim190$h)
on the MOS-SIS at the 3.6m CFHT (Le F\`evre et al 1995)
and DEEP2 took 90 nights ($\sim900$h) on DEIMOS at the 10m Keck telescope (Newman et al. 2012).

As a summary from these comparisons, the complementarity of deep galaxy spectroscopic surveys 
is evident in the parameter space defined by the number of spectra, area, depth, and redshift coverage. 
The VVDS survey is providing a unique galaxy
sample, selected in magnitude, with the widest redshift coverage of existing surveys
and an homogeneous treatment of spectra, 
and with depth, covered area, and number of spectra comparable to or better than other 
high redshift spectroscopic surveys existing today. 

\section{VVDS complete data release}
\label{release}

The data produced by the VVDS are all publicly released on a database with open access.
A complete information system has been developed with the data embedded in a 
reliable database environment, as described in Le Brun et al. (in preparation).
Queries combining the main survey parameters can be defined on the high--level user interface
or through SQL commands. Interactive plotting capabilities enable to identify the 
galaxies with redshifts on the deep images, and to view all the spectroscopic and
photometric information of each object with redshift, including spectra and thumbnails image extractions.
Spectra and images can be retreived in FITS format. 

The main parameters available on the database are as follows:
\begin{itemize}
\item Identification number, following the IAU notation ID-alpha-delta
\item $\alpha_{2000}, \delta_{2000}$ coordinates
\item Broad band photometric magnitudes u'grizJHKs
\item Spectroscopic redshift
\item Spectroscopic redshift reliability flag
\item Photometric redshift
\item TSR, SSR, PSR weights
\item Spectroscopic and photometric masks (region files)
\end{itemize}

In addition, specific parameters from connected surveys are available, depending 
on each of the VVDS fields, including e.g. Galex UV photometry, Spitzer-Swire photometry, VLA radio flux, etc.

The photometric data and spectra of 12\,430 Galactic stars are also available on this data release.

All the VVDS data can be queried on \texttt{http://cesam.lam.fr/vvds}.

\section{Summary}
\label{summary}

We have presented the VVDS surveys as completed, including the VVDS-Wide, VVDS-Deep,
VVDS-UltraDeep, and VVDS-LAE. In total the VVDS has measured spectroscopic redshifts 
for 34\,594 galaxies and $422$ AGN over a redshift range $0.007 \leq z_{spec} \leq 6.62$,
over an area up to 8.7 deg$^2$, for magnitude-limited surveys
with a depth down to $i_{AB}=24.75$ and 
line flux $F=1.5 \times 10^{18}$ erg/s/cm$^2$. 
The VVDS-Wide sample sums up to 25\,805 galaxies with $I_{AB} \leq 22.5$ over $0<z<2$, and a mean
spectroscopic redshift $\bar{z}=0.55$. The VVDS-Deep contains 11\,486 galaxies with 
$17.5 \leq I_{AB} \leq 24.0$ over $0<z<5$ and $\bar{z}=0.92$. The VVDS Ultra-Deep contains 
938 galaxies with measured redshifts with $23 \leq i_{AB} \leq 24.75$ over $0<z<4.5$ and  $\bar{z}=1.38$.
The VVDS-LAE sample adds 133 serendipitously discovered LAE to the high redshift populations with $2<z\leq6.62$.

Independent and deeper observations have been
obtained on $\sim1250$ galaxies, which has enabled to fully assess
the reliability of the redshift measurements. A reliability flag is associated
to each redshift measurement through inter-comparison of measurements
performed independently by two team members. We demonstrate that galaxies
with VVDS flags 2, 3, 4 and 9 have a reliability ranging from 83 to 100\%, making this the primary sample for science analysis. 
Galaxies with flag=1 can be used for  science analyses after taking
into account that they have a $\sim$50\% probability to be correct. We emphasize
that this probabilistic flag system enables a  robust statistical treatment
of the survey selection function, compiling a finer information 
than a simplistic good/bad redshift scheme. This leads to a well described selection
function with the TSR (Target Sampling Rate), PSR (Photometric
Sampling Rate), and SSR (Spectroscopic Success Rate) which are magnitude
and redshift dependent, as characterized in this paper, and available in our final data release.

We have emphasized the dependency of the 'redshift desert' on instrumental setups, 
and demonstrated that it can be successfully crossed when using a wavelength domain $3650 \leq \lambda \leq 9350$\AA. 
This wavelength range allows to follow the main emission/absorption
tracers like [OII]3727 or CaH\&K which would leave the domain for $z \geq 1.5$, 
while CIV-1549\AA ~and Ly$\alpha$-1215\AA ~would enter the domain
for $z \geq 1.32$ and $z \geq 2$ respectively. 

The basic properties of the VVDS samples have been described, including
the apparent and absolute magnitude distributions, as well as the ($M_u - M_r$, $M_r$) colour-magnitude
distribution showing a well defined colour bi-modality 
starting to be prominent at $1\leq z \leq 1.5$,
with a red sequence present up to $z\sim2$.
Averages of the observed VVDS-Deep and VVDS-UltraDeep spectra have been produced in redshift bins covering $0<z<5$.

The comparison of the VVDS survey with other spectroscopic redshift surveys
shows the unique place of the VVDS in the parameter space defined by the number of 
spectra, area, depth, and redshift coverage, complementary to other
surveys at similar redshifts.

All the VVDS data in this final release are made publicly available on a dedicated database 
available at \texttt{http://cesam.lam.fr/vvds}.

\begin{acknowledgements}
      We thank the referee, N. Padilla for his careful review of the manuscript and
      suggestions which have significantly improved this paper. We
      thank ESO staff for their continuous support for the VVDS surveys, particularly the
      Paranal staff conducting the observations and the ESO user support group in Garching.
      This work is supported by
      funding from the European Research Council Advanced Grant ERC-2010-AdG-268107-EARLY. 
      This survey is dedicated to the memory of Dr. Alain Mazure, who passed away when this
      paper was being refereed, and who has supported this work ever since its earliest stages.
\end{acknowledgements}

   \begin{table*}
      \caption[]{Comparison of VVDS surveys with other spectroscopic redshift surveys, by order of increasing mean redshift (updated from
                 Baldry et al. 2010)} 
      \[
         \begin{array}{lrlrclll}
           \hline \hline
            \noalign{\smallskip}
Survey         & Area    & Depth ~i_{AB}^a & N_{obj} &  z_{range} & z_{mean} & Selection & Reference \\ 
            \noalign{\smallskip}
            \hline
            \noalign{\smallskip}
SSRS2        &   5500  &  14.5  &  5\,369  &   0.0-0.08 &  0.03 & B<15.5        &  $da Costa et al. 1998$ \\
PSCz         &  34000  &  15.0  & 15\,411  &   0.0-0.1  &  0.03 & 60\mu_{AB}<9.5  &  $Saunders et al. 2000$ \\
2MRS         &  37000  &  15.2  & 23\,200  &   0.0-0.08 &  0.03 & K<15.2        &  $Erdo\u{g}du et al. 2006$ \\
SAPM         &   4300  &  15.7  &  1\,769  &   0.0-0.1  &  0.034 & BJ<17.1      &  $Loveday et al. 1992$ \\
6dFGS        &  17000  &  15.7  & 110\,256 &   0.0-0.1  &  0.05 & K<12.7        &  $Jones et al. 2009$ \\
CfA2         &  17000  &  14.5  &  13\,000 &   0.0-0.1  &  0.05 & B<15.5        &  $Falco et al. 1999$ \\
DURS         &   1500  &  15.6  &  2\,500  &   0.0-0.2  &  0.07 & BJ<17.0       &  $Ratcliffe et al. 1996$ \\
LCRS         &    700  &  17.5  & 26\,418  &   0.0-0.25 &  0.1  & R<17.5        &  $Shectman et al. 1996$ \\
2dFGRS       &   1500  &  18.0  & 250\,000 &   0.0-0.3  &  0.11 & BJ<19.4       &  $Colless et al. 2001$ \\
H-AAO        &    8.2  &  18.0  & 1\,056   &   0.0-0.55 &  0.14 & KAB<16.8      &  $Huang et al. 2003$ \\
ESP          &   23.3  &  18.0  & 3\,500   &   0.0-0.3  &  0.15 & BJ<19.4       &  $Vettolani et al. 1997$ \\
AGES         &    9.3  &  19.8  & 6\,500   &   0.0-0.5  &  0.2  & R<20 B_W<20.5 &  $Watson et al. 2009$ \\
GAMA         &    144  &  19.8  & 79\,599  &   0.0-0.6  &  0.25 & r<19.8, z<18.2, K_{AB}<17.6 & $Baldry et al. 2010$ \\
CNOC2        &    1.5  &  21.3  & 5\,000   &   0.12-0.55 & 0.3  & R<21.5        &  $Yee et al. 2000$ \\
Autofib      &    5.5  &  20.5  & 1\,700   &   0.0-0.75 &  0.3  & B_J<22         &  $Ellis et al. 1996$ \\
SDSS-LRG     &   8000  &  19.5  & 46\,748  &   0.16-0.47 & 0.3  & r<19.5 + CS^b & $Eisenstein et al. 2005$ \\
SDSS-MGS     &   9380  &  17.8  & 935\,000 &   0.0-0.6  &  0.3  & r<17.8, DR7   & $Abazajian et al. 2009$  \\
2SLAQ-LRG    &    180  &  20.3  & 11\,000  &   0.45-0.8 &  0.5  & i<19.8 + CS   & $Cannon et al. 2006$ \\
CFRS         &   0.14  &  22.5  & 600    &   0-1.5    & 0.55  & I_{AB} \leq 22.5 & $Lilly et al. 1995$ \\
zCosmos-Bright &    1.7  &  22.5  & 20\,000  &   0-1.5    &  0.55 & 17.5 \leq i_{AB} \leq 22.5 & $Lilly et al. 2007$  \\
{\bf VVDS-Wide}^c &   8.7 &  22.5  & 26\,178  &   0-1.5    & 0.55  & 17.5 \leq I_{AB} \leq 22.5  & $Garilli et al. 2008$  \\
PRIMUS       &    9.1  &  23.5  & 96\,599  &   0.0-1.2  &  0.6  & i_{AB} \leq 23  &  $Coil et al. 2011$ \\
WiggleZ      &   1000  &  21.0  & 100\,000 &   0.2-1.0  &  0.6  & NUV<22.8 + CS   &  $Blake et al. 2011$ \\
{\bf VVDS-Deep} & 0.74 &  24.00 & 11\,601  &   0-5      & 0.92  &  17.5 \leq I_{AB} \leq 24.0  & $Le F\`evre et al. 2005a, this paper$ \\
DEEP2        &    2.8  &  23.4  & 38\,000  &   0.7-1.4  &  1.0  & RAB<24.1 + CS &  $Newman et al. 2012$ \\
VIPERS       &      24 &  22.5  & 100\,000 &   0.5-1.5  &  1.0  & 17.5 \leq i_{AB} \leq 22.5 + CS & $Guzzo et al. 2013$ \\
{\bf VVDS-UDeep} &    0.14 &  24.75 & 941  &   0-4.5    & 1.38  &  23.0 \leq i_{AB} \leq 24.75 & $This paper$  \\
zCosmos-Deep &      1  &  23.75 & 2\,728   &   1.5-2.5  &  2.1  & B_{AB} \leq 25 + CS       & $Lilly et al. 2007$  \\
LBG-z3       &    0.38 &  24.8  & 1\,000   &   2.7-3.5  &  3.2  & R_{AB}<25.5 + CS          & $Steidel et al. 2003$ \\
VUDS         &      1  &  25    & 10\,000  &   2-6.7    &  3.7  & i_{AB}<25 + photo-z       & $Le F\`evre et al., in prep.$ \\ 
LBG-z4       &    0.38 &  25.0  & 300    &   3.5-4.5  &  4.0  & I_{AB}<25 +CS             &  $Steidel et al. 1999$ \\
\noalign{\smallskip}
\hline
\noalign{\smallskip}
{\bf VVDS-All}^d &  8.7  &  24.75 & 35\,016  &   0-5      & 1.2   & $Combined VVDS$           &  $This paper$ \\  
\noalign{\smallskip}
            \hline
         \end{array}
      \]
 \begin{list}{}{}
 \item[$^{\mathrm{a}}$] Equivalent depth in $i_{AB}$ using a flat spectrum transformation from the original survey depth.  
 \item[$^{\mathrm{b}}$] CS: Colour Selection, varies from survey to survey
 \item[$^{\mathrm{c}}$] The VVDS-Wide includes all objects with $17.5 \leq I_{AB} \leq 22.5$ in all 5 VVDS fields
 \item[$^{\mathrm{d}}$] Includes all objects in the 3 VVDS-Wide fields (1003+01, 1400+05 and 2217+00), in the VVDS-Deep fields 
(0226 and ECDFS), in the VVDS-UltraDeep field (0226-04), and the LAE emitters, as described in Table \ref{vvds}
 \end{list}
\label{comp_surveys}
   \end{table*}

\begin{thebibliography}{}

   \bibitem[2009]{aba} Abazajian et al., 2009, \apjs, 182, 543   

   \bibitem[2004]{abraham} Abraham, R., et al., 2004, \aj, 127, 2455   


   \bibitem[2005]{arnouts} Arnouts, S., et al., 2005, \apj, 619, 43   

   \bibitem[2007]{arnouts07} Arnouts, S., et al., 2007, \aap, 476, 137   

   \bibitem[2010]{baldry10} Baldry, I., et al., 2010, MNRAS, 404, 86  

   \bibitem[1996]{Sex} Bertin, E., Arnouts, S.,  1996, \aaps, 117, 393

   \bibitem[2009]{bielby} Bielby, R., et al., 2012, \aap, 545, 23    

   \bibitem[Blake et al. (2011)]{blake11} Blake, C., et al., 2011, \mnras, 418, 1707   

   \bibitem[2003]{bondi} Bondi, M., et al., 2003, \aap, 403, 857   

   \bibitem[2005]{bottini} Bottini, D., et al., 2005, \pasp, 117, 996   


   \bibitem[Cannon et al. (2006)]{cannon} Cannon, R., et al., 2006, \mnras, 372, 425   

   \bibitem[2011]{cassata11} Cassata, P., Le F\`evre, O., et al., 2011, \aap,  525, 143  

   \bibitem[2013]{cassata13} Cassata, P., Le F\`evre, O., et al., 2013, \aap, 556, 68  

   \bibitem[2008]{cimatti} Cimatti, A., et al., 2008, \aap, 482, 21     

   \bibitem[Coil et al. (2011)]{coil11} Coil, A., et al.,  2011, \apj, 741, 8   

   \bibitem[Colless et al. (1990)]{colless90} Colless, M., Ellis, R.S., Taylor, K., Hook, R.N., 1990, \mnras, 244, 408  

   \bibitem[Colless et al. (2001)]{colless} Colless, M., et al., 2001, \mnras, 328, 1039  

   \bibitem[Conti et al. 2001]{conti01}  Conti, G., et al., 2001, \pasp, 113, 452       

   \bibitem[Contini et al. (2012)]{contini} Contini, T., et al., 2012, \aap, 539, 91   

   \bibitem[Cucciati et al. 2012]{cucciati12} Cucciati, O., et al., 2012, \aap, 539, 31   

   \bibitem[Cucciati et al. 2010]{cucciati10} Cucciati, O., et al., 2010, \aap, 520A, 42  

   \bibitem[Cucciati et al. 2006]{cucciati06} Cucciati, O., et al., 2006, \aap, 458, 39   

   \bibitem[Cuillandre et al. 2012]{cuillandre12} Cuillandre, J.C., et al., 2012, Proc. SPIE 8448, Observatory Operations: Strategies, Processes, and Systems IV, 84480  

   \bibitem[1998]{dacosta} da Costa, L., et al., 1998, \aj, 116, 1   

   \bibitem[2004]{daddi} Daddi, E., et al., 2004, \apj, 617, 746   

   \bibitem[Davis et al., 2003]{davis} Davis, M., et al.,  2003, SPIE, 4834, 161    

   \bibitem[de la Torre et al., 2011]{de la Torre} de la Torre, S., et al., 2011, \aap, 525, 125  

   \bibitem[de Ravel et al., 2009]{deravel09} de Ravel, L., et al., 2009, \aap, 498, 379  


   \bibitem[Erdogdu et al. (2006)]{erdo} Erdo\u{g}du, P. et al., 2006, \mnras, 373, 45   

   \bibitem[Eisenstein et al. (2005)]{eisenstein} Eisenstein et al., 2005, \apj, 633, 560  

   \bibitem[Ellis et al. (1996)]{ellis} Ellis, R.S., et al. 1996, \mnras, 280, 235   

   \bibitem[Epinat et al. 2009]{epinat09} Epinat, B., et al., 2009, \aap, 504, 789   


   \bibitem[Falco et al. (1999)]{falco} Falco, E., et al.,1999, \pasp, 111, 438   

   \bibitem[2009]{fs09} Forster-Schreiber, N., et al., 2009, \apj, 706, 1364   

   \bibitem[Franzetti et al. (2007)]{franzetti} Franzetti, P., et al., 2007, A\&A, 465, 711  


   \bibitem[Garilli et al. (2008)]{garilli} Garilli, B., et al., 2008, \aap,  486, 683   

   \bibitem[Garilli et al. (2010)]{garilli2} Garilli, B., et al., 2010, PASP, 122, 827  

   \bibitem[Guzzo et al. (2008)]{guzzo08} Guzzo, L., et al., 2008, Nature, 451, 541  

   \bibitem[Guzzo et al. (2013)]{guzzo13} Guzzo, L., et al., 2013, arXiv:1303.2623  

   \bibitem[Hook et al. (2003)]{hook03} Hook, I., et al.,  2003, SPIE, 4841, 1645   

   \bibitem[Huang et al. (2003)]{huang} Huang et al. 2003, \apj, 584, 203   

   \bibitem[Ilbert et al. (2005)]{ilbert05} Ilbert, O., et al., 2005, \aap, 439, 863   

   \bibitem[Ilbert et al. (2006)]{ilbert06} Ilbert, O., et al., 2006, \aap, 457, 841   

   \bibitem[Ilbert et al. (2009)]{ilbert09} Ilbert, O., et al., 2009, \apj, 690, 1236  

   \bibitem[Iovino et al. 2005]{iovino} Iovino, A., et al., 2005, \aap, 442, 423   

   \bibitem[Jones et al. (2009)]{jones} Jones, D.H.,  et al., 2009, \mnras, 399, 683   

   \bibitem[Law et al. (2009)]{law09} Law, D., et al., 2009, \apj, 697, 2057    

   \bibitem[Lawrence et al. (2007)]{lawrence07} Lawrence, A. et al., 2007, MNRAS, 379, 1599   

   \bibitem[Le F\`evre et al. (1994)]{olf0} Le F\`evre, O., Crampton, D., Felenbok, P., Monnet, G., 1994, A\&A, 282, 325 

   \bibitem[Le F\`evre et al. 1995]{olf95} Le F\`evre, O., Crampton, D., Lilly, S.J., Hammer, F., Tresse, L., 1995, \apj, 455, 60  
 

   \bibitem[Le F\`evre et al. 2003]{olf03} Le F\`evre, O., et al., 2003, SPIE, 4841, 1670   

   \bibitem[Le F\`evre et al. (2004a)]{olfcdfs} Le F\`evre, O., et al., 2004a, A\&A, 428, 1043   

   \bibitem[Le F\`evre et al. (2004b)]{olfima} Le F\`evre, O., et al., 2004b, \aap, 417, 839  

   \bibitem[Le F\`evre et al. (2005a)]{olf2} Le F\`evre, O., et al., 2005a, \aap, 439, 845   

   \bibitem[Le F\`evre et al. (2005b)]{olf1} Le F\`evre, O., et al., 2005b, Nature, 437, 519  

   \bibitem[Le F\`evre et al. (2013)]{olf13} Le F\`evre, O., et al., 2013, \aap, arXiv:1307.6518  

   \bibitem[Lemoine-Busserolle et al. 2010]{lb10} Lemoine-Busserolle, M., Bunker, A., Lamareille, F., Kissler-Patig, M.,  2010, \mnras, 401, 1657  

   \bibitem[Lilly et al. (1991)]{lilly91} Lilly, S.J., Cowie, L.L., Gardner, J.P., 2001, \apj, 369, 79  

   \bibitem[Lilly et al. (1995)]{lilly1} Lilly, S.J., Le F\`evre, O., Crampton, D., Hammer, F., 1995a, \apj, 455, 50  
  

   \bibitem[Lilly et al. (1996)]{lilly3} Lilly, S. J., Le F\`evre, O., Hammer, F., Crampton, D., 1996, \apj, 460, 1  

   \bibitem[Lilly et al. (2007)]{lilly2} Lilly, S.J., Le F\`evre, O., et al., 2007, \apjs, 172, 70  

   \bibitem[2003]{lonsdale} Lonsdale, C.C., et al., 2003, \pasp, 115, 897   

   \bibitem[2011]{carlos} L\'opez-Sanjuan, C., et al.,  2011, \aap, 530, 20   

   \bibitem[Loveday et al. (1992)]{loveday} Loveday, J., et al., 1992, \apj, 390,338  

   \bibitem[Madau et al. (1996)]{mad} Madau, P., Ferguson, H.C.,  Dickinson, M.,  Giavalisco, M.,
    Steidel, C.C.,  Fruchter, A., 1996, \mnras, 283, 1388                   
  
   \bibitem[McCracken et al. 2003]{hjmcc} McCracken, H. J., Radovich, M., Bertin, E., Mellier, Y., 
    Dantel-Fort, M., Le F\`evre, O., Cuillandre, J. C., Gwyn, S., Foucaud, S., Zamorani, G., 2003, \aap, 410,17   

   \bibitem[Meneux et al. 2009]{meneux} Meneux, B., et al., 2009, \aap, 505, 463  
  
   \bibitem[Newman et al. (2012)]{deep2} Newman, J., et al., 2012, arXiv:1203.3192     

   \bibitem[Noll et al. (2004)]{noll} Noll, S., et al., 2004, \aap, 418, 885    

   \bibitem[Oliver et al. (2012)]{hermes} Oliver, S., et al., 2012, \mnras, 424, 1614   

   \bibitem[Ouchi et al. (2010)]{ouchi10} Ouchi, M., et al.,  2010, \apj, 723, 869   

   \bibitem[2004]{pierre} Pierre, M., et al., 2004, JCAP, 09, 011   

   \bibitem[Popesso et al. (2009)]{popesso} Popesso, et al.,  2009, \aap, 494, 443   

   \bibitem[Pozzetti et al. (2007)]{pozzetti07} Pozzetti, L., \aap, 474, 443 

   \bibitem[Radovich et al. (2004)]{radovich04} Radovich, M., et al., 2004, \aap, 417, 51   

   \bibitem[Ratcliffe et al. (1996)]{ratcliffe} Ratcliffe et al. 1996, \mnras, 281, 47   

   \bibitem[Saunders et al. (2000)]{saunders00} Saunders et al. 2000, \mnras, 317, 55   


   \bibitem[Shectman et al. (1996)]{Shectman96} Shectman, S.A., et al. 1996, \apj, 470, 172   

   \bibitem[2005]{sco} Scodeggio, M., et al., 2005, PASP, 117, 1284    

   \bibitem[Scoville et al. (2007)]{scov} Scoville, N., et al., 2007, \apjs, 172, 1  

   \bibitem[Steidel et al. 1996]{steidel96} Steidel, C.C., Giavalisco, M., Pettini, M., Dickinson, M., Adelberger, K.L., 1996, \apj, 462, 17    

   \bibitem[Steidel et al. (1999)]{steidel99} Steidel, C.C., Adelberger, K.L., Giavalisco, M., Dickinson, M., Pettini, M., 
   1999, \apj, 519, 1                     


   \bibitem[2003]{steidel03} Steidel, C.C., et al., 2003, \apj, 592, 728   

   \bibitem[Temporin et al. 2008]{temporin} Temporin, S., et al.,  2008, \aap, 482, 81   

   \bibitem[2007]{tresse} Tresse, L., et al., 2007, \aap, 472, 403  



   \bibitem[Vettolani et al. (1997)]{vettolani97} Vettolani, G., et al. 1997, \aap, 325, 954   

   \bibitem[Watson et al. (2009)]{watson} Watson, M.G., et al. 2009, \aap, 493, 339  

   \bibitem[Yee et al. (2000)]{Yee} Yee et al. 2000, \apjs, 129, 475   

   \bibitem[2000]{SDSS} York, D.G., et al.,  2000, AJ, 120, 1579  

   \bibitem[2006]{Zucca} Zucca, E., et al., 2006, \aap, 455, 879  

\end{thebibliography}
\end{document}